\newif\iffigs\figstrue
\documentclass[12pt]{article}
\usepackage{latexsym,amssymb}
\iffigs
             \input{epsf}
\else
\fi
\def\IC{\relax\,\hbox{$\inbar\kern-.3em{\rm C}$}}
\def\IG{\relax\,\hbox{$\inbar\kern-.3em{\rm G}$}}
\def\IB{\relax{\rm I\kern-.18em B}}
\def\ID{\relax{\rm I\kern-.18em D}}
\def\IL{\relax{\rm I\kern-.18em L}}
\def\IF{\relax{\rm I\kern-.18em F}}
\def\IH{\relax{\rm I\kern-.18em H}}
\def\II{\relax{\rm I\kern-.17em I}}
\def\IN{\relax{\rm I\kern-.18em N}}
\def\IP{\relax{\rm I\kern-.18em P}}
\def\IQ{\relax\,\hbox{$\inbar\kern-.3em{\rm Q}$}}
\def\bfzero{\relax\,\hbox{$\inbar\kern-.3em{\rm 0}$}}
\def\IK{\relax{\rm I\kern-.18em K}}
\def\IG{\relax\,\hbox{$\inbar\kern-.3em{\rm G}$}}
 \font\cmss=cmss10 \font\cmsss=cmss10 at 7pt
\def\IR{\relax{\rm I\kern-.18em R}}
\def\ZZ{\relax\ifmmode\mathchoice
{\hbox{\cmss Z\kern-.4em Z}}{\hbox{\cmss Z\kern-.4em Z}}
{\lower.9pt\hbox{\cmsss Z\kern-.4em Z}} {\lower1.2pt\hbox{\cmsss
Z\kern-.4em Z}}\else{\cmss Z\kern-.4em Z}\fi}
\def\bfone{\relax{\rm 1\kern-.35em 1}}

\def\inbar{\vrule height1.5ex width.4pt depth0pt}
\def\bfzero{\relax{\rm I\kern-.18em 0}}
\def\bfone{\relax{\rm 1\kern-.35em 1}}

\def\e{\epsilon}

\newcommand{\ft}[2]{{\textstyle\frac{#1}{#2}}}
\def\1bar{1\hskip -.275cm -}
\def\2bar{2\hskip -.275cm -}
\def\3bar{3\hskip -.275cm -}

\newsavebox{\uuunit}
\sbox{\uuunit}
                 {\setlength{\unitlength}{0.825em}
                      \begin{picture}(0.6,0.7)
                                      \thinlines
                                      \put(0,0){\line(1,0){0.5}}
                                      \put(0.15,0){\line(0,1){0.7}}
                                      \put(0.35,0){\line(0,1){0.8}}
                                     \multiput(0.3,0.8)(-0.04,-0.02){10}{\rule{0.5pt}{0.5pt}}
                      \end {picture}}

\makeatletter \@addtoreset{equation}{section} \makeatother
\newcommand{\be}{\begin{equation}}
\newcommand{\ee}{\end{equation}}
\newcommand{\ba}{\begin{eqnarray}}
\newcommand{\ea}{\end{eqnarray}}
\def\bfone{\relax{\rm 1\kern-.35em 1}}

\def\bfone{\relax{\rm 1\kern-.35em 1}}
\font\cmss=cmss10 \font\cmsss=cmss10 at 7pt
\begin{document}
\begin{titlepage}
\begin{flushright}
DFTT/05/33\\
\end{flushright}
\vskip 0.5cm
\begin{center}
{\LARGE {\bf M-theory FDA, Twisted Tori
}}\\[0.3cm]
{\LARGE {\bf and
}}\\[0.3cm]
{\LARGE {\bf
Chevalley Cohomology$^\dagger$}}\\[1.cm]
%{\LARGE {\bf    }}\\[1cm]
{\large Pietro Fr\'e$^*$}
{}~\\
\quad \\
{{\em  Dipartimento di Fisica Teorica, Universit\'a di Torino,}}
\\
{{\em $\&$ INFN - Sezione di Torino}}\\
{\em via P. Giuria 1, I-10125 Torino, Italy}~\quad\\
\end{center}
\begin{abstract}
The FDA algebras emerging from twisted tori compactifications of
M-theory with fluxes are discussed within the general classification
scheme provided by Sullivan's theorems and by Chevalley cohomology.
It is shown that the generalized Maurer Cartan equations which have appeared in
the literature, in spite of their complicated appearance, once suitably decoded
within cohomology, lead to trivial FDA.s, all new $p$--form generators
being contractible when  the $4$--form
flux is cohomologically trivial. Non trivial $D=4$ FDA.s can
emerge from non trivial fluxes but only if the cohomology class of the flux satisfies an
additional algebraic condition which appears  not to be satisfied in
general and has to be studied for each algebra separately.
As an illustration an exhaustive study of Chevalley cohomology for the simplest class of
SS algebras is presented but a general  formalism is developed, based on the structure of a double
elliptic complex,  which, besides providing the presented  results,  makes  possible the quick analysis of
compactification on any other twisted torus.
\end{abstract}
\vfill
\vspace{1.5cm}
\begin{flushleft}
{\em ~~~E-mail:}\\
{\em *) fre@to.infn.it}\\
\end{flushleft}
\vspace{2mm} \vfill \hrule width 3.cm {\footnotesize $^ \dagger $
This work is supported in part by the European Union RTN contract
MRTN-CT-2004-005104 and by the Italian Ministry of University (MIUR) under contract PRIN 2003023852}
\end{titlepage}
\section{Introduction}
\label{introsec}
Recently considerable attention has been devoted to flux
compactifications of String Theory and M--theory, since this provides
mechanisms to stabilize the moduli fields
\cite{fluxes0,fluxes1,fluxes2,oddstory,triveditripati,fluxesschulz,derendin}.
Within this context a particularly  interesting class of flux compactifications is
constituted by those on twisted tori. This is the contemporary understanding of the Scherk-Schwarz
\cite{Ssmecha} mechanism of mass generation from extra dimensions.
As it was explained by Hull and Reid-Edwards \cite{Hullo}, twisted
tori are just Lie group manifolds $\mathcal{G}$ modded by the action
of some discrete subgroup $\Delta \subset \mathcal{G}$ which makes
them compact.
\par
A general pattern only recently elucidated is the relation between
fluxes and gauge algebras. After dimensional reduction in presence of
flux compactifications  one ends up with some lower dimensional
gauged
supergravity and a particularly relevant question is that about the
structure of the gauge Lie algebra in relation with the choice of the
flux. Generalizing a concept originally introduced in
\cite{gaugedsugrapot1}, the authors of \cite{fluxgauge2} have
provided a very elegant and intrinsic classification of supergravity gaugings in terms of the so
called \textit{embedding matrix} and of its transformation properties
under the duality group $\mathrm{U}$. Later, in various papers, the relation between
the entries of the embedding matrix and the geometrical fluxes has
been elucidated for various cases of compactifications
\cite{fluxgauge3,fluxgauge1,lledolauramario}. Hence an
obvious question is that relative to the gauge algebra emerging from
flux compactifications on twisted tori. This question has been
addressed in a recent series of papers
\cite{GuidoSergioFDA,RiccaMarioSergioFDA1,RiccaGuidoSergioFDA,curvatureFDA1}
and it has been advocated that the gauge algebraic structures
emerging in $D=4$ supergravity from M--theory reduction on twisted
tori with fluxes do not fall in the class of Lie algebras $\mathbb{G}$ rather
have to be understood in the more general context of Free
Differential Algebras. A priori this is not too surprising since
M--theory itself, as all other higher dimensional supergravities, is
based on the gauging of an FDA. The algebraic structure that goes
under this name was independently discovered at the beginning of the eighties in Mathematics by
Sullivan \cite{sullivan} and in Physics by the present author in collaboration with R. D'Auria
\cite{fredauria}. Free Differential Algebras (FDA) are a  categorical extension of the
notion of Lie algebra and constitute the natural mathematical
environment for the description of the algebraic structure of higher
dimensional supergravity theory, hence also of string theory. The
reason is the ubiquitous presence in the spectrum of
string/supergravity theory of antisymmetric gauge fields ($p$--forms)
of rank greater than one. The very existence of FDA.s is a
consequence of Chevalley cohomology of ordinary Lie algebras and
Sullivan has provided us with a very elegant classification scheme of
these algebras based on two structural theorems rooted in the set up
of such an elliptic complex.
\par
As I already noted in a paper of about two decades ago
\cite{comments},  FDA.s have the additional fascinating property that,
differently from ordinary Lie algebras they already encompass their
own gauging. Indeed the first of Sullivan's structural theorems, which is
in some sense analogous to Levi's theorem for Lie algebras, states
that the most general FDA is a semidirect sum of a so called minimal
algebra $\mathbb{M}$ with a contractible one $\mathbb{C}$. The
generators of the minimal algebra are physically interpreted as the
connections or \textit{potentials}, while the contractible generators
are physically interpreted as the \textit{curvatures}. The real
hard--core of the FDA is the minimal algebra and it is obtained by
setting the contractible generators (the curvatures) to zero. The
structure of the minimal algebra $\mathbb{M}$, on its turn, is beautifully
determined by Chevalley cohomology of $\mathbb{G}$. This happens to
be the content of Sullivan's second structural theorem.
\par
In the present paper my goal is that of recasting the findings   and
the  equations presented in
\cite{GuidoSergioFDA,RiccaMarioSergioFDA1,RiccaGuidoSergioFDA,curvatureFDA1}
into the general scheme of FDA.s. In particular I want to unveil the
cohomological interpretation of those equations and answer the basic
question, namely what is the structure of the minimal algebra $\mathbb{M}$.
My result is  that notwithstanding their apparent complicated appearance, when
appropriately decoded within the language of
cohomology, the zero curvature  Maurer Cartan equations
presented in \cite{RiccaGuidoSergioFDA,curvatureFDA1} actually define
a trivial minimal algebra $\mathbb{M}=\mathbb{G}$, all higher degree
generators being contractible. This is necessarily the case if the $4$--form flux
is cohomologically trivial. Non trivial $4$--form fluxes might instead lead
to non trivial $D=4$ minimal algebras. However this  happens if and only if a  certain algebraic relation
is satisfied by the cohomology class to which the flux is assigned. I
will spell out explicitly such an algebraic condition showing that it is
not generically satisfied by simple non degenerate SS algebras. It can be, instead,  satisfied
for degenerate algebras.
\par
In this paper  my main goal is to establish a compact, index free, formalism
to deal with twisted tori compactifications which unveils the
underlying mathematical set up,  that of a double elliptic
complex. Indeed the basic objects we deal with are  space--time $p$--forms valued in
Chevalley $q$--cochains and we have two boundary operators, one acting on the base space
and one on the Chevalley fibre. Once reformulated in this way the
spectrum of generators of the $D=4$ FDA is immediately read off from
the Chevalley cohomology groups of the fibre Lie algebra
$\mathbb{G}$. The generalized structure constants of the FDA are also
elegantly rewritten in terms of the Poincar\'e pairing form on the
Chevalley complex and the question whether the minimal algebra is
trivial or not can also be answered within this framework.
\par
The paper is organized as follows.
\par
 In section \ref{chevalley} I
summarize the set up of Chevalley cohomology. In subsections
\ref{pairing}, \ref{Sscohomo} and \ref{orthogosec} I calculate the
cohomology groups of SS algebras and construct an explicit basis for
their cochain spaces.
\par
In section \ref{sullivano} I recall the general set up of FDA.s and
I summarize Sullivan structural theorems.
In subsection \ref{minimSS} I discuss the non trivial FDA.s
associated with SS algebras.
\par
In section \ref{superFDA} I illustrate the working of Sullivan's
theorem with the example of the super FDA of M--theory.
\par
In section \ref{Mtwisted} I elaborate the cohomological analysis of
twisted tori compactifications and I set up the double elliptic
complex formalism. Then in subsection \ref{zerocurve} I single out
the cohomological interpretation of the zero curvature equations
showing that they always lead to a trivial minimal algebra, all
$p$--form generators being contractible. In the last section
\ref{introcurve} I introduce the curvature and I discuss the bearing
of cohomologically non trivial fluxes.
\par
Finally section \ref{concludo} contains my conclusions.
\section{Chevalley Cohomology}
\label{chevalley}
As a necessary preparatory step for my discussion let me shortly
recall the setup of the Chevalley elliptic complex leading to Lie
algebra cohomology. This will also fix my notations and conventions.
\par
Let us consider a (super) Lie algebra $\mathbb{G}$ identified through its
structure constants $\tau^{I}_{\phantom{I}JK}$ which are
alternatively introduced through the commutation relation of the generators\footnote{
In view of my main goal which is the analysis of FDA.s emerging in bosonic M-theory compactifications
I adopt a pure Lie algebra notation. Yet every definition presented here has a straightforward
extension to superalgebras and indeed when I will recall the discussion of how the FDA of
 M-theory emerges from the application of
Sullivan structural theorems is within the scope of super Lie algebra  cohomology}
\begin{equation}
  \left[ T_I \, , T_K\right] \, = \, \tau^{I}_{\phantom{I}JK} \, T_I
\label{commrel}
\end{equation}
or the Cartan Maurer equations:
\begin{equation}
  \partial e^I = \ft 12 \, \tau^{I}_{\phantom{I}JK} \, e^J \wedge e^K
\label{MC1}
\end{equation}
where $e^I$ is an abstract set of left--invariant $1$--forms. The
isomorphism between the two descriptions (\ref{commrel}) and
(\ref{MC1}) of the Lie algebra is provided by the duality relations:
\begin{equation}
  e^I(T_J) = \delta^I_J
\label{dualityrele}
\end{equation}
A $p$-cochain $\Omega^{[p]}$ of the Chevalley complex is just an exterior $p$-form
on the Lie algebra with constant coefficients, namely:
\begin{equation}
  \Omega^{[p]} = \Omega_{I_1\dots I_p} \, e^{I_1} \, \wedge \, \dots
  \wedge \, e^{I_p}
\label{pcochain}
\end{equation}
where the antisymmetric tensor $\Omega_{I_1\dots I_p} \in
\bigwedge^{p} \,\mbox{adj} \mathbb{G}$ which belongs to the $p$-th
antisymmetric power of the adjoint representation of $\mathbb{G}$ has constant
components. Using the Maurer cartan equations (\ref{MC1}) the
coboundary operator $\partial$ has a pure algebraic action on the
Chevalley cochains:
\begin{eqnarray}
  \partial \, \Omega^{[p]} & = & \partial \, \Omega_{I_1 \dots
  I_{p+1}} \, e^{I_1} \, \wedge \, \dots \, \wedge \, e^{I_{p+1}}
  \nonumber\\
\partial \, \Omega_{I_1 \dots
  I_{p+1}} & = & (-)^{p-1} \, \frac{p}{2} \,\, \tau^{R}_{\phantom{R}[I_1I_2}
  \, \Omega_{I_1 \dots
  I_{p+1}]R}
\label{domeg}
\end{eqnarray}
and Jacobi identities guarantee the nilpotency of this operation
$\partial^2=0$. The cohomology groups $H^{[p]}\left(\mathbb{G} \right)
$ are constructed in standard way. The $p$--cocycles
$\Omega^{[p]}$ are the closed forms $ \partial \Omega^{[p]} =0$ while
the exact $p$-forms  or $p$-coboundaries  are those $\Lambda^{[p]}$  such that they can be
written as $\Lambda^{[p]} = \partial \Phi^{[p-1]}$ for some suitable
$p-1$-forms $\Phi^{[p-1]}$. The $p$-th cohomology groups is spanned by
the $p$--cocycles modulo the $p$--coboundaries. Calling $C^{p}\left(
\mathbb{G}\right)$ the linear space of $p$-chains the operator
$\partial$ defined in eq.(\ref{domeg}) induces a sequence of linear maps $\partial_{p}$:
\begin{equation}
  C^{0}\left(
\mathbb{G}\right) \, \stackrel{\partial_0}{\Longrightarrow} \,  C^{1}\left(
\mathbb{G}\right) \, \, \stackrel{\partial_1}{\Longrightarrow} \, C^{2}\left(
\mathbb{G}\right) \, \, \stackrel{\partial_2}{\Longrightarrow} \, C^{3}\left(
\mathbb{G}\right) \, \, \stackrel{\partial_3}{\Longrightarrow} \, C^{4}\left(
\mathbb{G}\right) \, \, \stackrel{\partial_4}{\Longrightarrow} \,
\dots
\label{sequence}
\end{equation}
and we can summarize the definition of the Chevalley cohomology groups in
the standard form used for all elliptic complexes:
\begin{equation}
  H^{(p)}\left( \mathbb{G}\right)  \equiv \frac{\mbox{ker}  \, \partial_p}{\mbox{Im}\, \partial_{p-1}}
\label{Hgroups}
\end{equation}
\par
On the Chevalley complex it is also convenient to introduce the
operation of contraction with a tangent vector and then of Lie
derivative.
\par
The contraction operator $i_X$ associates to every tangent vector,
namely to every element of the Lie algebra  $X \in \mathbb{G}$ a
linear map from the space $C^{p}\left( \mathbb{G}\right) $ of the $p$-cochains to the space
$C^{p-1}\left( \mathbb{G}\right)$ of the $p-1$-cochains:
\begin{equation}
  \forall \, X \, \in \, \mathbb{G}\quad ; \quad i_X \, : \; C^{p}\left( \mathbb{G}\right)
  \, \mapsto \, C^{p-1}\left( \mathbb{G}\right)
\label{idiX}
\end{equation}
Explicitly we set:
\begin{equation}
  \forall X = X^M \, T_M \, \in \, \mathbb{G} \quad ; \quad i_X \,
  \Omega^{[p]} = p \, X^M \, \Omega_{MI_1 \dots I_{p-1}} \, e^{I_1}
  \, \wedge \, \dots \, \wedge  \, e^{I_{p-1}}
\label{idXdefi}
\end{equation}
Next I introduce the Lie derivative $\ell$ which to every element of
the Lie algebra $X \in \mathbb{G}$ associates a map from the space of
$p$--cochains into itself:
\begin{equation}
  \forall X = X^M \, T_M \, \in \, \mathbb{G} \quad ; \quad \ell_X \, : \; C^{p}\left( \mathbb{G}\right)
  \, \mapsto \, C^{p}\left( \mathbb{G}\right)
\label{ellXmap}
\end{equation}
The map $\ell_X$ is defined as follows:
\begin{equation}
  \ell_X \, \equiv \, i_X \, \circ \, \partial + \partial \, \circ \,
  i_X
\label{ellXdefi}
\end{equation}
and satisfies the necessary property in order to be a representation
of the Lie algebra:
\begin{equation}
 \left [ \ell_X, \ell_Y \right
 ] \, = \, \ell_{[X,Y]}
\label{oppone}
\end{equation}
By explicit calculation we find in components:
\begin{equation}
  \ell_X \, \Omega^{[p]} = (-)^{p-1} \, p \, X^M \,
  \tau^{R}_{\phantom{R}M[I_1} \, \Omega_{I_2I_3 \dots I_p]R} \,
  e^{I_1} \, \wedge \, \dots \, \wedge \, e^{I_p}
\label{ellXincompo}
\end{equation}
Furthermore if $X$ and $Y$ are any two $\mathbb{G}$ Lie algebra--valued space--time forms, respectively of
degree $x$ and $y$, by direct use of the above definitions, you can
easily verify the following identity which holds true on any
$p$--cochain $\mathcal{C}^{[p]}$:
\begin{equation}
 \left(  i_X \circ \ell_Y + (-)^{xy+1} \ell_Y \right) \, \mathcal{C}^{[p]} =  - i_{\left
 [ X \, , \, Y \right ]} \, \mathcal{C}^{[p]}
\label{LxiY}
\end{equation}
and which will of great help in my following analysis.
%%%%%%%%%%%%%%%%%%%%%%%%%%%%%%%%%%%%%%%%%%%%%%%%%
\subsection{The pairing form and Poincar\'e duality on the Chevalley
complex}
\label{pairing}
Chevalley cochains are antisymmetric tensors on a finite dimensional
vector space, namely the Lie algebra $\mathbb{G}$. For this reason
the space ${C}^{p}$ is a finite dimensional
and can in fact be decomposed into subspaces of predictable
dimension. Let us call
\begin{equation}
  d = \mbox{dim} \, \mathbb{G}
\label{dequaldimG}
\end{equation}
Then the vector space of all cochains ${C}$ has dimension $2^d$ and we
can write:
\begin{equation}
\begin{array}{cccccccccccc}
  {C} & = & {C}^0 & \oplus & {C}^1 & \oplus & {C}^2 &\oplus & \dots & \oplus & {C}^d \\
  2^d & = & 1 & + & \left(\begin{array}{c}
    d \\
    1 \
  \end{array} \right) & + & \left(\begin{array}{c}
    d \\
    2 \
  \end{array} \right)  & + & \dots & + & \left(\begin{array}{c}
    d \\
    d \
  \end{array} \right)
\end{array}
\label{pongone}
\end{equation}
Each finite dimensional $p$-cochain space can now be decomposed
into the following triplet of subspaces:
\begin{equation}
  {C}^n = \mathbf{\Gamma}^{[n]} \, \oplus \, \partial \, \mathbf{\Xi}^{[n-1]} \, \oplus \,
  \mathbf{\Xi}^{[n]}
\label{decompino}
\end{equation}
where, by definition we have posed:
\begin{eqnarray}
  \partial \, \mathbf{\Xi}^{[n-1]} & \equiv & \mbox{Im}\,\partial_{n-1}\nonumber\\
\mathbf{\Gamma}^{[p]} & \equiv &\mbox{orthogonal complement of $\mbox{Im}\,\partial_{n-1}$ in
$\mbox{ker}\,\partial_n$}\nonumber\\
\mathbf{\Xi}^{[n]} & \equiv &\mbox{orthogonal complement of $\mbox{ker}\,\partial_n$ in
${C}^n$}\nonumber\\
\label{spaccopere}
\end{eqnarray}
We give a name to the dimensions of these subspaces:
\begin{equation}
  \begin{array}{rclcrclcrcl}
    \mbox{dim}\,\mathbf{\Gamma}^{[n]}  & \equiv & h_n & ; & \mbox{dim}\,\partial \mathbf{\Xi}^{[n-1]} & \equiv & \wp_n & ; &
    \mbox{dim}\,\mathbf{\Xi}^{[n]} & \equiv & r_n \\
  \end{array}
\label{nomini}
\end{equation}
and we have the obvious sum rules:
\begin{equation}
  h_n +\wp_n+ r_n = \left(\begin{array}{c}
    d \\
    n \
  \end{array} \right)
\label{summarulla}
\end{equation}
\paragraph{Volume preserving algebras} From now on I restrict my
attention to Lie algebras $\mathbb{G}$ that are characterized by the
additional property:
\begin{equation}
  \tau^{M}_{\phantom{M}MI} = 0
\label{capelki}
\end{equation}
imposed on the structure constants (see
\cite{RiccaMarioSergioFDA1,RiccaGuidoSergioFDA,curvatureFDA1}).
For such algebras we immediately prove the following lemma. $\forall
X \in \mathbb{G}$:
\begin{equation}
  \ell_X \, \mathrm{Vol} = 0
\label{invariantvolume}
\end{equation}
where
\begin{equation}
  \mathrm{Vol} = \frac{1}{d!} \, \epsilon_{I_1, I_2 ,\dots , I_{d}}
  \, e^{I_1} \wedge \, \dots \, \wedge \,e^{I_d}
\label{volume_form}
\end{equation}
is the the volume $d$-form. Then
the orthogonal decomposition can be better formulated by introducing
a pairing form on the space cochains. This will lead to Poincar\'e
duality. Let me consider two forms of complementary degree $\Omega^{[p]}$ and
$\Psi^{[d-p]}$. I define their scalar product by setting:
\begin{equation}
  \Omega^{[p]} \, \wedge \,\Psi^{[d-p]} \, \equiv \, \langle \,
  \Omega^{[p]} \, , \, \Psi^{[d-p]}  \rangle \, \mbox{Vol}
\label{cicici}
\end{equation}
Next I can prove a very important property of the pairing form I have
just introduced which holds true for volume preserving algebras, namely under the additional condition
(\ref{capelki}).
To this effect note that for such subalgebras for which eq.
(\ref{capelki}) is true, any arbitrary $d-1$ form is closed:
\begin{equation}
  \forall \, \Omega^{[d-1]} \in C^{d-1} \; : \, \partial \, \Omega^{[d-1]} =0
\label{tuttochiusoin6}
\end{equation}
The proof is elementary. The components of $\partial \, \Omega^{[d-1]}$
are $\tau^{M}_{[I_1 I_2} \, \Omega_{I_3I_4 \dots I_{d-1}]M}$ from which
it follows that the index $M$ being different from $I_3I_4 \dots
I_{d-1}$ is necessarily equal to either $I_1$ or $I_2$ and this makes
the contribution zero in view of eq. (\ref{capelki}).
Secondly I note that:
\begin{eqnarray}
0 & = & \partial \left( \Omega^{[p-1]} \, \wedge \, \Psi^{[d-p]} \right)  \nonumber\\
\null & = & \partial \Omega^{[p-1]} \,\wedge \, \Psi^{[d-p]} \, + \,
(-)^{p-1} \, \Omega^{[p-1]} \,\wedge \,  \partial \Psi^{[d-p]}
\nonumber \\
\null &=& \left( \langle \partial \Omega^{[p-1]} \, , \, \Psi^{[d-p]}\rangle +
(-)^{p-1} \, \langle \Omega^{[p-1]} \,,  \,  \partial \Psi^{[d-p]}
\rangle \right )* \mbox{Vol}
\label{adjunt}
\end{eqnarray}
Hence I conclude that we have the very important formula realizing
Poincar\'e duality:
\begin{equation}
  \langle \partial \Omega^{[p-1]} \, , \, \Psi^{[p]}\rangle \, = \, (-)^{p-1} \,
  \langle \Omega^{[p-1]} \,,  \,  \partial \Psi^{[p]}
\rangle
\label{octopusPoinca}
\end{equation}
Equation (\ref{octopusPoinca}) can be used to derive recursion
relation that determine the values of the numbers $\wp_n$ and $r_n$ in
terms of the formal Hodge numbers $h_n$. The first observation is
that, thanks to (\ref{octopusPoinca}) the subspace $\partial \,\mathbf{\Xi}^{[d-n-1]}$,
namely $\mbox{Im}\, \partial_{d-n-1}$ is orthogonal to the space of
$n$--cycles, \emph{i.e.} to $\mathbf{\Gamma}^{[n]}\oplus \partial \, \mathbf{\Xi}^{[n-1]}$:
\begin{eqnarray}
  \langle \,\mathbf{\Gamma}^{[n]}\oplus \partial \, \mathbf{\Xi}^{[n-1]} \, , \, \partial \, \mathbf{\Xi}^{[d-n-1]}
  \rangle & = &   \langle \,\partial \left( \mathbf{\Gamma}^{[n]}\oplus \partial \, \mathbf{\Xi}^{[n-1]} \right) \, , \,  \mathbf{\Xi}^{[d-n-1]}
  \rangle\nonumber\\
  &=&0
\label{perpequa}
\end{eqnarray}
The above equation implies that:
\begin{equation}
  \wp_{d-n} = r_n =\left( \begin{array}{c}
    d \\
    n \
  \end{array}\right) \, - \, h_n \, - \, \wp_n
\label{primarelaziawp}
\end{equation}
Applying now the same relation in the opposite direction, namely
exchanging $(d-n) \leftrightarrow n$, we obtain:
\begin{equation}
  \wp_{n} =\left( \begin{array}{c}
    d \\
    d-n \
  \end{array}\right) \, - \, h_{d-n} \, - \, \wp_{d-n}
\label{secondarelaziawp}
\end{equation}
and summing together eq. (\ref{primarelaziawp}) with eq.
(\ref{secondarelaziawp}) we obtain the standard Poincar\'e duality on
cohomology groups:
\begin{equation}
  h_{d-n} = h_n
\label{poincadulae}
\end{equation}
On the other hand, from their very definition we can obtain a recursion
relation on the numbers $\wp_n$:
\begin{equation}
  \wp_n = \left(\begin{array}{c}
    d \\
    n-1 \
  \end{array} \right) - h_{n-1} -\wp_{n-1}
\label{recursiarela}
\end{equation}
which can be solved, once combined with eq.s (\ref{poincadulae}) and
(\ref{primarelaziawp}).
In table \ref{poincartab} we exhibit the complete table of the
relevant numbers $h_n,\wp_n,r_n$ for a $7$--dimensional Lie algebra,
which will be the case of interest for our subsequent study of
M--theory compactification on the so called twisted tori.
\begin{table}
  \centering
  \begin{eqnarray*}
  \begin{array}{||c|c|c|c|c||}
  \hline
  \hline
    n & h_n & \wp_n & r_n & h_n+\wp_n+r_n = \left(\begin{array}{c}
      7 \\
      n \\
    \end{array} \right)  \\
    \hline
    \hline
   7 &h_7= h_0 = 1 & 0 & 0 & 1 \\
   6 & h_6 = h_1 & 7-h_1 & 0 & 7 \\
   5 & h_5 =h_2 & 14-h_2 + h_1 & 7-h_1 & 21 \\
   4 & h_4= h_3 & 21-h_3 + h_2 -h_1 & 14-h_2 + h_1 & 35 \\
   3 & h_3 & 14-h_2 + h_1 & 21-h_3 + h_2 - h_1 & 35 \\
   2 & h_2 & 7-h_1 & 14-h_2+h_1 & 21 \\
   1 & h_1 & 0 & 7-h_1 & 7 \\
   0 & h_0=1   & 0 & 0     & 1 \\
    \hline
  \end{array}
   &&\nonumber\\
   \null &&\nonumber\\
\end{eqnarray*}
  \caption{Decomposition of Chevalley cochain spaces for a $7$-dimensional Lie algebra $\mathbb{G}$.
  As one sees the entire table is parametrized by three independent numbers $h_{3,2,1}$,
  the dimensions of the cohomology groups
  $H^{1,2,3}(\mathbb{G})$,respectively.
  \label{poincartab}}
\end{table}
From table (\ref{poincartab}) we realize that the following identity
also follows, from the above relations, in full generality:
\begin{equation}
  \wp_n = r_{n-1}
\label{optime}
\end{equation}
%%%%%%%%%%%%%%%%%%%%%%%%%%%%%
Two other important formal properties of the pairing form are the
following ones. Let $W \in \mathbb{G} $ be a space--time $1$-form
valued element of the Lie algebra. Considering M--theory
compactifications on twisted tori we will see the role of such an
object. At the level of the present mathematical discussion $W$ is
just a Lie algebra element that also anticommutes with the Chevalley
$1$--forms $e^I$. This being the case we have:
\begin{equation}
  \langle \Psi^{[d-p+1]} \, , \, i_W \, \Omega^{[p]} \rangle =
  \langle i_W \,\Psi^{[d-p+1]} \, , \,   \Omega^{[p]} \rangle
\label{iwmigra}
\end{equation}
Using the pairing form and relying on its properties, I can now
construct a complete orthonormal basis for the space of Chevalley
chains. I introduce the following index conventions.
In degree $p \le \frac{d}{2}$ we introduce two type of indices. The
unbarred ones
\begin{equation}
  \alpha_{(p)} \, = \, 1,...,h_p=h_{d-p}
\label{greekindeces}
\end{equation}
which enumerate the cohomology classes and will be attributed to an
orthonormal basis of forms for the spaces $\mathbf{\Gamma}^{[p]}$:
\begin{equation}
  \langle {\Gamma}^{[p]}_{\alpha_{(p)}} \, , \, {\Gamma}^{[d-p]}_{\beta_{(p)}}
  \rangle = \delta_{\alpha_{(p)}\beta_{(p)}}
\label{Gammabassa}
\end{equation}
and the barred ones:
\begin{equation}
  \bar{\alpha}_{(p)} \, = \, 1,...,r_p=r_{d-p-1}
\label{greekindeces2}
\end{equation}
which enumerate the basis elements for the space $\mathbf{\Xi}^{[p]}$,
namely the space of those $p$--cochains whose derivative is strictly
non zero $\partial \mathbf{\Xi}^{[p]} \ne 0\, $:
\begin{equation}
  \langle {\Xi}^{[p]}_{\bar{\alpha}_{(p)}} \, , \, \partial \, {\Xi}^{[d-p-1]}_{\bar{\beta}_{(p)}}
  \rangle = \delta_{\bar{\alpha}_{(p)}\bar{\beta}_{(p)}}
\label{Xibassa}
\end{equation}
Furthermore we  always have  the following orthogonality relations:
\begin{eqnarray}
\langle \mathbf{\Gamma}^{[p]} \, , \, \partial \, \mathbf{\Xi}^{[d-p-1]} \rangle & = &
0 \nonumber\\
\langle \mathbf{\Gamma^{[p]}} \, , \, \mathbf{\Xi}^{[d-p]} \rangle & = &
0 \nonumber\\
\langle \mathbf{\Xi}^{[p]} \, , \,  \mathbf{\Xi}^{[d-p]} \rangle & = &
0 \nonumber\\
\langle \partial \, \mathbf{\Xi}^{[p]} \, , \,  \partial \, \mathbf{\Xi}^{[d-p]}
\rangle & = & 0
\label{ortogonalia}
\end{eqnarray}
which are true by definition of the spaces $\mathbf{\Xi}^{[p]}$ and
$\mathbf{\Gamma}^{[p]}$.
In the case of a $7$--dimensional Lie algebra $\mathbb{G}$, which is that
relevant for twisted tori compactifications, in order to avoid a too
clumsy notation I distinguish among the various $\alpha_{(p)}$ and
$\bar{\alpha}_{(p)}$ indices by changing alphabet type:  since the
relevant values of $p$ are just three this is possible.
Hence by definition I set:
\begin{eqnarray}
\alpha_{(1)},\beta_{(1)}\dots & \equiv & \alpha,\beta \, = \, 1, \dots , h_1 \nonumber\\
\bar{\alpha}_{(1)},\bar{\beta}_{(1)}\dots & \equiv & \bar{\alpha},\bar{\beta} \, = \, 1, \dots ,7- h_1 \nonumber\\
\alpha_{(2)},\beta_{(2)}\dots & \equiv & a,b \dots\, = \, 1, \dots , h_2 \nonumber\\
\bar{\alpha}_{(2)},\bar{\beta}_{(2)}\dots & \equiv & \bar{a},\bar{b} \dots\, = \, 1, \dots ,14- h_2+h_1 \nonumber\\
\alpha_{(3)},\beta_{(3)}\dots & \equiv & x,y \dots\, = \, 1, \dots , h_3 \nonumber\\
\bar{\alpha}_{(3)},\bar{\beta}_{(3)}\dots & \equiv & \bar{x},\bar{y} \dots\, = \, 1, \dots ,21-h_3+ h_2-h_1 \nonumber\\
\label{omninusindex}
\end{eqnarray}
Correspondingly I can construct a complete orthonormal basis of cochains
as it is displayed in table \ref{unciricamo}:
\begin{table}
  \centering
\begin{eqnarray*}
  \begin{array}{||ccc|c|l||}
  \hline
    {\Gamma}_\alpha^{[1]} & \leftrightarrow & {\Gamma}_\alpha^{[6]} & \langle {\Gamma}_\alpha^{[1]} \, ,\,
    {\Gamma}_\beta^{[6]}\rangle = \delta_{\alpha\beta} & \alpha,\beta = 1,\dots , h_1 \\
    \hline
   {\Xi}_{\bar{\alpha}}^{[1]} & \leftrightarrow & {\Xi}_{\bar{\alpha}}^{[5]} & \langle {\Xi}_{\bar{\alpha}}^{[1]} \, ,
   \,
   \partial{\Xi}_{\bar{\alpha}}^{[5]}\rangle = \delta_{\bar{\alpha}
   \bar{\beta}} & \bar{\alpha},\bar{\beta} = 1,\dots , 7-h_1 \\
  \hline
    {\Gamma}_a^{[2]} & \leftrightarrow & {\Gamma}_a^{[5]} & \langle {\Gamma}_a^{[2]} \, ,\,
    {\Gamma}_b^{[5]}\rangle = \delta_{ab} & a,b = 1,\dots , h_2 \\
    \hline
   {\Xi}_{\bar{a}}^{[2]} & \leftrightarrow & {\Xi}_{\bar{a}}^{[4]} & \langle {\Xi}_{\bar{a}}^{[2]} \, ,\,
   \partial\, {\Xi}_{\bar{b}}^{[4]}\rangle =
   \delta_{\bar{a}
   \bar{b}} & \bar{a},\bar{b} = 1,\dots , 14-h_2+h_1 \\
   \hline
    {\Gamma}_x^{[3]} & \leftrightarrow & {\Gamma}_x^{[4]} & \langle {\Gamma}_x^{[2]} \, ,\,
    {\Gamma}_y^{[5]}\rangle = \delta_{xy} & x,y = 1,\dots , h_3 \\
    \hline
  {\Xi}_{\bar{x}}^{[3]} & \leftrightarrow & {\Xi}_{\bar{x}}^{[3]} & \langle {\Xi}_{\bar{x}}^{[3]} \, ,\,
   \partial {\Xi}_{\bar{y}}^{[3]}\rangle =
   \delta_{\bar{x}
   \bar{y}} & \bar{x},\bar{y} = 1,\dots , 21-h_3+h_2 -h_1\\
   \hline
   \hline
  \end{array}
\end{eqnarray*}
 \caption{Construction of a well adapted basis of $n$--cochains for a $7$--dimensional Lie algebra.
  \label{unciricamo}}
\end{table}
\paragraph{Some additional useful general relations}
There are still some general relations that I can prove and that will
be of great help in my subsequent analysis of Free Differential
Algebras of M-theory. From the definition of pairing (\ref{cicici})
we easily conclude that for volume preserving algebras
(\ref{capelki})  and $\Psi^{[p]}$, $\Omega^{[d-p]}$ any two cochains
of complementary degree we have:
\begin{equation}
\langle \ell_W \, \Psi^{[p]} \, , \, \Omega^{[d-p]} \rangle = (-)^p
\langle  \, \Psi^{[p]} \, , \, \ell_W \Omega^{[d-p]} \rangle
\label{spostoellW}
\end{equation}
 From the very definition of Lie derivative
(\ref{ellXdefi}), it follows that:
\begin{equation}
  \langle \ell_W \mathbf{\Gamma}^{(p)} \, , \,
  \mathbf{\Gamma}^{(d-p)} \rangle = 0
\label{primaorto}
\end{equation}
Indeed we have $ \ell_W \mathbf{\Gamma}^{(p)}  = \partial \, i_W \,
\mathbf{\Gamma}^{(p)}$ since $\partial \, \mathbf{\Gamma}^{(p)} = 0$
and $\langle \partial \, i_W \,
\mathbf{\Gamma}^{(p)} \, , \,  \mathbf{\Gamma}^{(d-p)} \rangle$ = $\langle i_W \,
\mathbf{\Gamma}^{(p)} \, , \, \partial \, \mathbf{\Gamma}^{(d-p)}
\rangle$
= $0 $. For the same reason we also have:
\begin{equation}
  \langle \ell_W \mathbf{\Gamma}^{(p)} \, , \,
 \partial \mathbf{\Xi}^{(d-p-1)} \rangle = 0
\label{seconorto}
\end{equation}
This orthogonality relations implies that:
\begin{equation}
  \ell_W \mathbf{\Gamma}^{(p)} \, \subset \, \partial \, \mathbf{\Xi}^{(p-1)}
\label{dentroli!}
\end{equation}
\par
Similarly  we have :
\begin{eqnarray}
\ell_W  \mathbf{\Xi}^{(p)} & \subset & \mathbf{\Gamma}^{(p)}\, \oplus \, \mathbf{\Xi}^{(p)} \, \oplus \, \partial \,
  \mathbf{\Xi}^{(p-1)}\nonumber\\
  \ell_W \partial \mathbf{\Xi}^{(p-1)} & \subset & \partial \,
  \mathbf{\Xi}^{(p-1)}
\label{dentroqui!}
\end{eqnarray}
All together eq.s (\ref{dentroli!}) and (\ref{dentroqui!}) imply
that the representation of the algebra $\mathbb{G}$ on the space of
$p$--cochains is always upper triangular:
\begin{equation}
  \ell_W \, \left(\begin{array}{c}
    \mathbf{\Xi}^{(p)} \\
 \mathbf{\Gamma}^{(p)}    \\
  \partial \,  \mathbf{\Xi}^{(p-1)}  \
  \end{array} \right) \, = \, \left( \begin{array}{c|c|c}
    \star & \star & \star \\
    \hline
    0 & 0 & \star \\
    \hline
    0 & 0 & \star \
  \end{array}\right) \, \, \left(\begin{array}{c}
    \mathbf{\Xi}^{(p)} \\
 \mathbf{\Gamma}^{(p)}    \\
   \partial \,  \mathbf{\Xi}^{(p-1)} \
  \end{array} \right)
\label{ornato}
\end{equation}
\par
\paragraph{Decomposition of $p$-cochains along the three subspaces}
\par
Once we have have constructed an orthogonal basis, each Chevalley
$p$--cochain can be decomposed along it and we can define its
projections onto the three subspaces $\mathbf{\Gamma}^{(p)}$, $\partial \,
\mathbf{\Xi}^{(p-1)}$ and $\mathbf{\Xi}^{(p-1)}$.
\par
Let $\Omega^{(p)}$ be a generic $p$--cochain. I can set:
\begin{eqnarray}
\Omega^{(p)} & = & P^{(p)}_\perp \left [\Omega \right ] \,\oplus \, \partial \, Q^{(p-1)} \left [\Omega \right ] \, \oplus \,
P^{(p)}_\| \left [\Omega \right ] \nonumber\\
\label{decompo3}
\end{eqnarray}
where
\begin{equation}
  P^{(p)}_\perp \left [\Omega^{(p)} \right ] \, \in \,
  \mathbf{\Gamma}^{(p)} \quad ; \quad \partial \, Q^{(p-1)} \left [\Omega^{(p)} \right ] \, \in \,
 \partial \, \mathbf{\Xi}^{(p-1)} \quad ; \quad P^{(p)}_\| \left [\Omega^{(p)} \right ] \, \in \,
  \mathbf{\Xi}^{(p)}
\label{treproiezioni}
\end{equation}
The projection onto the three orthogonal subspaces can be
performed using the pairing form and the orthogonal basis.
Explicitly I can write:
\begin{eqnarray}
 P^{(p)}_\perp \left [\Omega^{(p)} \right ]  & = & \sum_{\alpha_{(p)}=1}^{h_p} \,\langle \Omega^{(p)} \, , \,\Gamma^{(d-p)}_{\alpha_{(p)}} \rangle \,
 \Gamma^{(p)}_{\alpha_{(p)}} \nonumber\\
 P^{(p)}_\| \left [\Omega^{(p)} \right ]  & = & \sum_{\bar{\alpha}_{(p)}=1}^{r_p} \,\langle \Omega^{(p)} \,
 , \,\partial \Xi^{(d-p-1)}_{\bar{\alpha}_{(p)}} \rangle \,
 \Xi^{(p)}_{\bar{\alpha}_{(p)}} \nonumber\\
  Q^{(p-1)}_\| \left [\Omega^{(p)} \right ]  & = & \sum_{\bar{\alpha}_{(p)}=1}^{r_{p-1}} \,\langle \Omega^{(p)} \,
  , \,\Xi^{(d-p)}_{\bar{\alpha}_{(p-1)}} \rangle \,
 \Xi^{(p-1)}_{\bar{\alpha}_{(p-1)}}
\label{projettori}
\end{eqnarray}
The above formulae will be very useful in my analysis of the FDA
emerging from M--theory compactifications on twisted tori
%%%%%%%%%%%%%%%%%%%%%%%%%%%%%%%%%%%%%%%%%%%%%%%%%
\subsection{Calculation of the relevant cohomology groups for SS algebras}
\label{Sscohomo}
As a concrete illustration of the general scheme I consider the
example of Scherk Schwarz algebras introduced in \cite{RiccaGuidoSergioFDA} and already used as a toy model in
\cite{RiccaMarioSergioFDA1,curvatureFDA1}. The algebra $\mathbb{G}$
underlying such an example is identified by the following choice of
the structure constants:
\begin{eqnarray}
  I &=& 0,i \quad; \quad  i=1,\dots , 6 \nonumber\\
  \tau^{I}_{\phantom{I}JK} & = &
  \left \{
  \begin{array}{ll}
   \tau^{i}_{\phantom{i}0 k}  = &  T^i_{\phantom{i}j}  = \mbox{antisymmetric matrix} \\
    0 & \mbox{otherwise}  \
  \end{array} \right.
\label{tauspeciale}
\end{eqnarray}
This means that the Maurer Cartan equations (\ref{MC1}) take the
following specific form:
\begin{eqnarray}
\partial e^{0} & = & 0 \nonumber\\
\partial e^{i} & = & e^0 \, \wedge \, T^i_{\phantom{i}j} \, e^j
\label{Gexemplo}
\end{eqnarray}
The $6 \times 6$ antisymmetric matrix $T^i_{\phantom{i}j}$ can always be skewed
diagonalized by means of automorphisms of the algebra:
\begin{eqnarray}
  T^i_{\phantom{i}j} & = & \left( \begin{array}{c|c|c}
    m_1  \mathbf{\epsilon} & \mathbf{0_{2\times 2}} & \mathbf{0_{2\times 2}} \\
    \hline
   \mathbf{0_{2\times 2}} & m_2  \mathbf{\epsilon} & \mathbf{0_{2\times 2}} \\
   \hline
   \mathbf{ 0_{2\times 2}} & \mathbf{0_{2\times 2}} & m_3  \mathbf{\epsilon} \
  \end{array}\right) \nonumber\\
\mathbf{\epsilon} & = & \left( \begin{array}{cc}
  0 & 1 \\
  -1 & 0
\end{array}\right)
\label{Tskewdiag}
\end{eqnarray}
and the Maurer Cartan equations (\ref{Gexemplo}) turn out to define
seven different non isomorphic algebras, depending on the eigenvalue
pattern structure. From our previous discussion we learned that there
are three relevant Chevalley Cohomology groups $\mathrm{H}^{4,2,1}(\mathbb{G})$. In
this section I calculate them for the algebras (\ref{Gexemplo}).
\par
\paragraph{Computation of $\mathrm{H}^{2}(\mathbb{G})$}
\par I begin with the second cohomology group. I want to compute
its dimension and the structure of the tensors which correspond to
its elements. The first task is to find the $2$--cocycles
\begin{equation}
  \partial \, \Sigma^{[2]} \, = \, 0 \quad \Leftrightarrow \quad
  \tau^{R}_{[IJ} \, \Sigma_{K]R} = 0
\label{2cocycles}
\end{equation}
Using the specific form (\ref{tauspeciale}) of the structure
constants we reduce condition (\ref{2cocycles}) to the form
$\tau^{r}_{[IJ} \, \Sigma_{K]r} = 0$ and we have two possibilities for the free indices,
either $IJK = ijk$ or $IJK=0jk$. In the first case, due to
(\ref{tauspeciale}), the equation is automatically satisfied and there
is no condition on the tensor $\Sigma$. For the second choice of free
indices I get instead the following condition:
\begin{eqnarray}
 0 & = & \tau^{r}_{[0j} \, \Sigma_{k]r}  = \ft 1 3 \left(  \tau^{r}_{0j} \,
 \Sigma_{kr}\,  + \, \tau^{r}_{jk} \,
 \Sigma_{0r} \, + \, \tau^{r}_{k0} \,
 \Sigma_{jr} \right ) \nonumber\\
 & = & \ft 1 3 \left[ \left( T\cdot \sigma\right) _{jk} +  \left( T\cdot \sigma\right) _{kj} \right] \\
 \label{2cocycondo}
\end{eqnarray}
where I have named $\sigma$ the antisymmetric matrix $\Sigma_{rs}$.
Due to antisymmetry of both matrices the last line of equation (\ref{2cocycondo}) can
be read as
\begin{equation}
  \left[ T \, , \, \sigma\right] = 0
\label{TcommutaSig}
\end{equation}
Hence the space of $2$--cocycles is spanned by the $6$  tensors with
arbitrary components $\Sigma_{0r}$ (no-condition on them) plus the solutions of
eq.(\ref{TcommutaSig}). Let us study its meaning.
\par
Being antisymmetric both $\sigma$ and $T$ are elements of the $\mathrm{SO(6)}$
Lie algebra. Once $T$ is skew diagonalized it belongs to the Cartan
subalgebra of $\mathrm{SO(6)}$ and I can identify the skew eigenvalues $m_i$
with its components along a basis of Cartan generators:
\begin{equation}
  T = {\rm i} \, \left( m_1 \mathcal{H}^1 \, + \, m_2 \mathcal{H}^2 \, + \, m_3
  \mathcal{H}^3\right)\, \in \, \mathrm{CSA} \subset \mathrm{SO(6)}
\label{TinCartan}
\end{equation}
Let me now parametrize the most general form that $\sigma$ can have:
\begin{eqnarray}
  \sigma & = & {\rm i} \, \left(f_i \mathcal{H}^i \, \right) + x_\alpha
  \, X^\alpha \, + \, y_\alpha \, Y^\alpha \nonumber\\
X^\alpha & = & \left(E^\alpha - E^{-\alpha} \right) \nonumber\\
Y^\alpha & = & {\rm i} \, \left(E^\alpha + E^{-\alpha} \right)
\label{parametrosigma}
\end{eqnarray}
where $\alpha \in \Delta_+(D_3)$ are the positive roots of the $D_3$
simple algebra, $E^{\pm\alpha}$ the step-up (step-down) operators
associated to each root and $\mathcal{H}_i$ its Cartan generators.
All Cartan generators commute with a Cartan element by definition,
hence the $3$--parameters $f_i$ are free. On the other hand we have:
\begin{eqnarray}
\left[ T \, , \, X^\alpha \right]  & = & \alpha(T) \, Y^\alpha \nonumber\\
\left[ T \, , \, Y^\alpha \right]  & = & - \alpha(T) \, X^\alpha
\label{commuTXY}
\end{eqnarray}
Hence for each positive root $\alpha$ orthogonal to the vector
$\{m_1,m_2,m_3\}$ which identifies $T$ there is a pair $(x_\alpha \, , \,  y_\alpha)$ of additional
parameters in $\sigma$, besides the three $f_i$. Let me recall the
form of positive roots for the $D_3$ root systems. They are six and
precisely $\epsilon_i \pm \epsilon_j$, having denoted by $\epsilon_i$
an orthonormal basis in $\mathbb{R}^3$. Hence I can write:
\begin{equation}
  \begin{array}{ccccccc}
    \alpha_1(T) & = & m_1 -m_2 & ; & \alpha_2(T) & = & m_1 + m_2 \\
    \alpha_3(T) & = & m_1 - m_3 & ; & \alpha_4(T) & = & m_1+m_3 \\
    \alpha_5(T) & = & m_2-m_3 & ; & \alpha_6(T) & = & m_2 +m_3 \
  \end{array}
\label{all&alpha}
\end{equation}
I can now discuss the number of  roots orthogonal to $T$, in relation
with the eigenvalue structure. If $m_i$ are all non vanishing and
different from each other then no $\alpha_i(T)$ vanishes.
Correspondingly the number of parameters in $\sigma$ is just three
and the dimension of the $2$-cocycle space is
$\mathbf{dim}[\mbox{ker}\,\partial_2]
= 6+3=9$, where $6$ are the tensors $\Sigma_{0r}$ and $3$
the skew diagonal $\sigma$.s. On the other hand if the $m_i$ are all
non zero but two are equal among themselves, say $m_1=m_2 \ne m_3 $, it is
evident that there is just one root orthogonal to $T$. In
this case the dimension of the $2$-cocycle space grows. We have
$\mathbf{dim}[\mbox{ker}\,\partial_2] = 6+3+ 2 \times 1 =11$.
If all eigenvalues $m_i$ are equal and non zero, $m_1=m_2=m_3$, there
are three roots orthogonal to $T$ and we have
$\mathbf{dim}[\mbox{ker}\,\partial_2] = 6+3+ 2 \times 3 =15$. It
should be noted that the same conclusion holds true if the
eigenvalues are equal in absolute value but opposite in sign. This
discussion explains the first three entries in the second column of
table \ref{cohomotab}.
\begin{table}
  \centering
  \begin{eqnarray*}
  \begin{array}{||l||r|c|c||c|c||r|c|c||}
  \hline
  \hline
    \mathbb{G} & \mbox{dim\,ker} \partial_2 & \mbox{dim\,Im} \partial_1 & \mathbf{h_2} & \mbox{dim\,ker}
    \partial_1
     & \mathbf{h_1} & \mbox{dim\,ker} \partial_4 & \mbox{dim\,Im} \partial_3 & \mathbf{h_4} \\
    \hline
    \hline
    m_1\ne m_2  & 9=6+3 & 6  & \mathbf{3} & 1 & \mathbf{1} & 23=20+3 & 20 & \mathbf{3} \\
     \ne m_3 & \null & \null & \null & \null & \null & \null & \null &
    \null \\
    \hline
    m_1 = m_2  & 11=6+3 & 6  & \mathbf{5} & 1 & \mathbf{1} & 25=20+3 & 20 &
    \mathbf{5} \\
    \ne m_3 & +2 \times 1 & \null & \null & \null & \null & + 2\times 1 & \null & \null \\
   \hline
   m_1 = m_2  & 15=6+3 & 6  & \mathbf{9} & 1 & \mathbf{1} & 29=20+3 & 20 &
    \mathbf{9} \\
    = m_3 & +2 \times 3 & \null & \null & \null & \null & + 2\times 3 & \null & \null \\
   \hline
   m_1 \ne m_2 \, ,  & 9=6+3 & 4  & \mathbf{5} & 1+2 & \mathbf{3} & 23=20+3 & 16 &
    \mathbf{7} \\
    m_3=0 & \null & \null & \null & \null & \null & \null & \null &
    \null \\
   \hline
   m_1 = m_2\,  & 11=6+3 & 4  & \mathbf{7} & 1+2 & \mathbf{3} & 25=20+3 & 12 &
    \mathbf{13} \\
     m_3=0 & +2 \times 1 & \null & \null & \null & \null & + 2\times 1 & \null & \null \\
   \hline
   m_1 \ne 0 \, , & 13=6+3 & 2  & \mathbf{11} & 1+4 & \mathbf{5} & 27=20+3 & 12 &
    \mathbf{15} \\
    m_2= m_3=0 & +2 \times 2 & \null & \null & \null & \null & + 2\times 2 & \null & \null \\
   \hline
   m_1 = m_2=  & 21=6+3 & 0  & \mathbf{21} & 1+6 & \mathbf{7} & 35=20+3 & 0 &
    \mathbf{35} \\
    m_3=0 & +2 \times 6 & \null & \null & \null & \null & + 2\times 6 & \null & \null \\
   \hline
   \hline
  \end{array} &&\nonumber\\
   \null &&\nonumber\\
\end{eqnarray*}
  \caption{Chevalley Cohomology groups of the Lie algebra defined by the Maurer Cartan equations
  in eq.(\ref{Gexemplo}). As one sees the cohomology crucially depends on the eigenvalue structure
  of the $6 \times 6$ matrix $T$.  \label{cohomotab}}
\end{table}
Before addressing the remaining cases where one or more of the skew
eigenvalues $m_i$ vanish, let me discuss the image of the
$\partial_1$ map in the cohomology sequence (\ref{sequence}). This
space is spanned by the trivial $2$--cocycles of the form:
\begin{equation}
  \Sigma^{[2]} \, \in \, \mathrm{Im} \partial_1 \, \Rightarrow \,
  \Sigma^{[2]} = \partial U^{[1]}
\label{2boundaries}
\end{equation}
In components this means tensors of the form
\begin{equation}
  \tau^{I}_{JK} \, U_I \, \Rightarrow \, \tau^{i}_{0k} \, U_i =
  T^{i}_{\phantom{i}j} \, U_i
\label{tau&U}
\end{equation}
where $U_i$ is an arbitrary $6$-vector. If the matrix $T$ is
non degenerate all tensors $\Sigma_{0r}$ can be put in such a form: it
suffices to pose $U_i = (T^{-1})_i^j \Sigma_{0j}$. Hence when no
eigenvalue $m_i$ vanishes, we have $
\mathbf{dim}[\mathrm{Im}\partial_1]=6$ and this explains also the
first three entries in the $4$-th column of table \ref{cohomotab}.
Indeed we always have:
\begin{equation}
  \mathbf{dim}[\mathrm{H}^p] = \mathbf{dim}[\mathrm{ker }\partial_p]
  - \mathbf{dim}[\mathrm{Im }\partial_{p-1}]
\label{ovviadim}
\end{equation}
On the contrary if some skew eigenvalue vanish the dimension of $\mathbf{dim}[\mathrm{Im}\partial_1]$ decreases
since $T$ projects onto a smaller space. In particular if one eigenvalue vanishes, say $m_3=0$ we have
$\mathbf{dim}[\mathrm{Im}\partial_1]=4$, if two vanish $m_2=m_3=0$ we get $\mathbf{dim}[\mathrm{Im}\partial_1]=2$
and when all vanish we have $\mathbf{dim}[\mathrm{Im}\partial_1]=0$.
\par
Keeping this information in mind let me return to discuss the space
of cocycles in the degenerate case when some eigenvalues vanish. The
first case to consider is $m_1 \ne m_2 \ne 0, m_3=0$. For these
eigenvalues no root is orthogonal to $T$ hence the space of
$2$--cocycles has dimension  $\mathbf{dim}[\mbox{ker}\,\partial_2] =
6+3=9$. The decreased dimension of the $2$--boundary space explains
the increase in the number of cohomology classes, $h_2 = 5$ (see
table \ref{cohomotab}). If $m_3=0$ and $m_1=m_2$ are equal, there is
just one root orthogonal to $T$ and we obtain $\mathbf{dim}[\mbox{ker}\,\partial_2] =
6+3+2=11$ which leads to $h_2=7$. When two eigenvalues vanish
$m_2=m_3$ there are two roots orthogonal to $T$ and we get $\mathbf{dim}[\mbox{ker}\,\partial_2] =
6+3+2\times 2=13$, leading to $h_2=11$. Finally when all eigenvalues vanish
all six roots are orthogonal to $T$ and we obtain  $\mathbf{dim}[\mbox{ker}\,\partial_2] =
6+3+2\times 6=21$ leading to $h_2=21$. This completes the explanation
of the second, third and fourth columns of table \ref{cohomotab},
namely the computation of $h_2$ for all the seven different algebras encoded
in eq.(\ref{Gexemplo}).
\par
\paragraph{Computation of $\mathrm{H}^{1}(\mathbb{G})$} This is
fairly simple. A $1$-form $\Sigma^{[1]}$ is just a vector in the
adjoint representation $\Sigma_I$ and the cocycle condition is:
\begin{equation}
  \tau^F_{\phantom{F}IJ} \, \Sigma_F = 0
\label{1cociclo}
\end{equation}
As we did above we have to distinguish two cases in the choice of the
free indices $IJ$. Either $IJ = ij$ or $IJ =0j$. In the first case we
obtain no condition on the vector $\Sigma_I$. In the second case we
obtain:
\begin{equation}
 0 =  \tau^{f}_{0j} \, \Sigma_i = T_{j}^{\phantom{j}i} \, \Sigma_i
\label{condo1ciclo}
\end{equation}
As long as the matrix $T$ is non degenerate it does not admit null
eigenvectors and hence the space of $1$--cocycles is
$1$--dimensional, it just spanned by the vectors of the form:
$\Sigma_0 \ne 0 \, , \, \Sigma_i = 0$. This explains the first three
entries in the fifth column of table \ref{cohomotab}. Then one easily
realizes that for each vanishing skew eigenvalue $m_i=0$ we have a
pair of null eigenvectors and this explains the remaining entries in
the fifth column. The fact that the sixth column coincides with the fifth,
namely that each $1$-cocycle is also a cohomology class follows from the
fact that there are no $1$--coboundaries, since in Chevalley cohomology we cannot produce a
$1$--cochain starting from a $0$--chain, that is just from a constant
number.
This concludes the calculation of $\mathrm{H}^{1}(\mathbb{G})$
\par
\paragraph{Computation of $\mathrm{H}^{4}(\mathbb{G})$} We are
interested both in $\mathrm{H}^{3}(\mathbb{G})$ and in
$\mathrm{H}^{4}(\mathbb{G})$, but by Poincar\'e duality these cohomology groups are isomorphic and it suffices
to calculate one of the two. It turns out that
$\mathrm{H}^{4}(\mathbb{G})$ is easier. A $4$-cochain $\Sigma^{[4]}$
is represented by a $4$--index tensor $\Sigma_{IJKL}$ and the cocycle
condition is:
\begin{equation}
  0 = \tau^{R}_{\phantom{R}[IJ } \, \Sigma_{KLM]R}
\label{4ciclocondo}
\end{equation}
Once again we have to distinguish two cases in the choice of the free
indices $IJKLM$, either $ijk\ell m$ or $0jk\ell m$. In the first case
the cocycle condition is identically satisfied and we get no
restriction on the components of $\Sigma^{[4]}$. In the second case
the equation splits again in two sectors:
\begin{equation}
  \begin{array}{ccc}
    \begin{array}{ccc}
      \tau^{r}_{\phantom{r}[0i}\, \Sigma_{k\ell m]r} & = & 0 \\
    \end{array} & \Rightarrow & \begin{array}{ccc}
      \Sigma_{0 k\ell m}  & = & \mbox{free} \\
    \end{array} \\
    \Downarrow & \null & \null \\
     \begin{array}{ccc}
      \tau^{r}_{\phantom{r}0[i}\, \Sigma_{k\ell m]r} & = & 0 \\
    \end{array} & \Rightarrow & T^{\phantom{i}r}_{[i} \, \Sigma_{k\ell m]r} = 0 \
  \end{array}
\label{quadraequa}
\end{equation}
Hence we have to analyze the equation in the right-bottom corner of
eq.(\ref{quadraequa}). $\Sigma_{k\ell mr}$ is a rank $4$
antisymmetric tensor in $d=6$. Hence we can write it as:
\begin{equation}
  \Sigma_{k\ell mr} = \epsilon_{k\ell mr u v} \, \sigma^{uv}
\label{epsilonsigma}
\end{equation}
where $\sigma^{uv}$ is a rank $2$ antisymmetric tensor, namely a
matrix. In this way we can translate eq.(\ref{quadraequa}) as
follows:
\begin{eqnarray}
0 & = & \epsilon^{abijk\ell m} \, T^{\phantom{i}r}_{i} \, \Sigma_{k\ell m r} \nonumber\\
\null & = &  \epsilon^{abijk\ell m} \, \epsilon_{k\ell mr uv} \,
T^{\phantom{i}r}_i \, \sigma^{uv} \nonumber\\
\null & \Downarrow & \null \nonumber\\
0 & = & \delta^{ab i}_{ruv} \, T^{\phantom{i}r}_{i} \, \sigma^{uv}
\label{rettangolo}
\end{eqnarray}
Since $T^{\phantom{i}r}_{i}$ is antisymmetric, there is no
contribution from $i=r$, what remains is
\begin{equation}
0 =  T^{\phantom{i}a}_{v} \, \sigma^{bv}\,  - \, T^{\phantom{i}b}_{v} \, \sigma^{av}
\label{lastessaequa}
\end{equation}
This is the same equation (see (\ref{TcommutaSig})) we have already
solved in the calculation of $\mathrm{H}^{2}(\mathbb{G})$. Hence in
the various cases the dimensions of $ \mbox{ker} \, \partial_4$ are
\begin{equation}
  \mbox{dim} \, ker \, \partial_4 = \underbrace{20}_{\Sigma_{0k\ell m}}
   \, + \, \mbox{$\#$ of solutions of $\left[ T \, , \, \sigma \right] = 0$}
\label{numsolvo}
\end{equation}
This explains all the entries in the seventh column of table
(\ref{cohomotab}). It remains to calculate the dimension of
$\mbox{Im} \, \partial_3$. To this effect we argue in the following
way. The space $ \partial \mathbf{\Xi}^{[3]}$ is spanned by tensors of
the form:
\begin{equation}
  \Sigma_{0ijk} = T^{\phantom{a}r}_{[i} \, U_{jk]r}
\label{pettodipollo}
\end{equation}
where $U_{jkr}$ is an arbitrary rank $3$ antisymmetric tensor in
$d=6$. Hence the operation described by equation (\ref{pettodipollo})
defines a linear map from the space of rank $3$ antisymmetric tensors into itself:
\begin{equation}
  \mathcal{T} \quad  : \quad \bigwedge^3 \mathbb{R} \, \,\rightarrow \, \, \bigwedge^3 \mathbb{R}
\label{testone}
\end{equation}
It suffices to derive the $20 \times 20$ matrix $\mathcal{T} $ and
calculate its rank, namely the number of non--vanishing eigenvalues,
depending on the eigenvalue structure of $T$. If we choose a
lexicographic order for the independent components of a rank $3$
antisymmetric tensor, $i.e.$ $U_{123},U_{124},\dots ,U_{456}$, the
explicit form of the matrix $\mathcal{T}$ is given below:
{\tiny
\begin{eqnarray}
&\mathcal{T}=&\nonumber\\
& \left(  \matrix{ 0 & {m_
    2} & 0 & 0 & 0 & 0 & 0 & 0 & 0 & 0 & 0 & 0 & 0 & 0 & 0 & 0 &
   0 & 0 & 0 & 0 \cr
    -{m_2} & 0 & 0 & 0 & 0 & 0 & 0 & 0 & 0 & 0 & 0 & 0 & 0 & 0 &
   0 & 0 & 0 & 0 & 0 & 0 \cr 0 & 0 & 0 & {m_
    3} & 0 & 0 & 0 & 0 & 0 & 0 & 0 & 0 & 0 & 0 & 0 & 0 & 0 & 0 &
   0 & 0 \cr 0 & 0 &
    -{m_3} & 0 & 0 & 0 & 0 & 0 & 0 & 0 & 0 & 0 & 0 & 0 & 0 & 0 &
   0 & 0 & 0 & 0 \cr 0 & 0 & 0 & 0 & 0 & 0 & 0 & 0 & 0 & 0 & {m_
    1} & 0 & 0 & 0 & 0 & 0 & 0 & 0 & 0 & 0 \cr 0 & 0 & 0 & 0 & 0 &
   0 & {m_3} & {m_2} & 0 & 0 & 0 & {m_
    1} & 0 & 0 & 0 & 0 & 0 & 0 & 0 & 0 \cr 0 & 0 & 0 & 0 & 0 &
    -{m_3} & 0 & 0 & {m_2} & 0 & 0 & 0 & {m_
    1} & 0 & 0 & 0 & 0 & 0 & 0 & 0 \cr 0 & 0 & 0 & 0 & 0 &
    -{m_2} & 0 & 0 & {m_3} & 0 & 0 & 0 & 0 & {m_
    1} & 0 & 0 & 0 & 0 & 0 & 0 \cr 0 & 0 & 0 & 0 & 0 & 0 &
    -{m_2} & -{m_3} & 0 & 0 & 0 & 0 & 0 & 0 & {m_
    1} & 0 & 0 & 0 & 0 & 0 \cr 0 & 0 & 0 & 0 & 0 & 0 & 0 & 0 & 0 &
   0 & 0 & 0 & 0 & 0 & 0 & {m_
    1} & 0 & 0 & 0 & 0 \cr 0 & 0 & 0 & 0 &
    -{m_1} & 0 & 0 & 0 & 0 & 0 & 0 & 0 & 0 & 0 & 0 & 0 & 0 & 0 &
   0 & 0 \cr 0 & 0 & 0 & 0 & 0 &
    -{m_1} & 0 & 0 & 0 & 0 & 0 & 0 & {m_3} & {m_
    2} & 0 & 0 & 0 & 0 & 0 & 0 \cr 0 & 0 & 0 & 0 & 0 & 0 &
    -{m_1} & 0 & 0 & 0 & 0 & -{m_3} & 0 & 0 & {m_
    2} & 0 & 0 & 0 & 0 & 0 \cr 0 & 0 & 0 & 0 & 0 & 0 & 0 &
    -{m_1} & 0 & 0 & 0 & -{m_2} & 0 & 0 & {m_
    3} & 0 & 0 & 0 & 0 & 0 \cr 0 & 0 & 0 & 0 & 0 & 0 & 0 & 0 &
    -{m_1} & 0 & 0 & 0 & -{m_2} &
    -{m_3} & 0 & 0 & 0 & 0 & 0 & 0 \cr 0 & 0 & 0 & 0 & 0 & 0 & 0 &
   0 & 0 &
    -{m_1} & 0 & 0 & 0 & 0 & 0 & 0 & 0 & 0 & 0 & 0 \cr 0 & 0 & 0 &
   0 & 0 & 0 & 0 & 0 & 0 & 0 & 0 & 0 & 0 & 0 & 0 & 0 & 0 & {m_
    3} & 0 & 0 \cr 0 & 0 & 0 & 0 & 0 & 0 & 0 & 0 & 0 & 0 & 0 & 0 &
   0 & 0 & 0 & 0 &
    -{m_3} & 0 & 0 & 0 \cr 0 & 0 & 0 & 0 & 0 & 0 & 0 & 0 & 0 & 0 &
   0 & 0 & 0 & 0 & 0 & 0 & 0 & 0 & 0 & {m_
    2} \cr 0 & 0 & 0 & 0 & 0 & 0 & 0 & 0 & 0 & 0 & 0 & 0 & 0 & 0 &
   0 & 0 & 0 & 0 & -{m_2} & 0 \cr  }\right)& \nonumber\\
\label{Taumatrone}
\end{eqnarray}}
and calculating its determinant we find:
\begin{equation}
  \mbox{Det} \, \mathcal{T} = {{m_1}}^4\,{{m_2}}^4\,{{m_3}}^4\,
  {\left( {{m_1}}^4 + {\left( {{m_2}}^2 - {{m_3}}^2 \right) }^2 -
      2\,{{m_1}}^2\,\left( {{m_2}}^2 + {{m_3}}^2 \right)  \right) }^2
\label{determi}
\end{equation}
Furthermore the $10$ independent skew eigenvalues of $\mathcal{T}$ (which is
antisymmetric) are:
\begin{equation}
  \begin{array}{ccc}
    \lambda_1 & = & m_1 \\
    \lambda_2 & = & -m_1 \\
    \lambda_3 & = & m_2 \\
    \lambda_4 & = & -m_2  \\
    \lambda_5 & = & m_3  \\
    \lambda_6 & = & - m_3 \\
    \lambda_7 & = & m_1 + m_2 + m_3 \\
    \lambda_8 & = & -m_1 + m_2 + m_3\\
    \lambda_9 & = & m_1 - m_2 + m_3 \\
    \lambda_{10} & = & m_1 + m_2 - m_3 \
  \end{array}
\label{eigenvalli}
\end{equation}
Inspection of eq.(\ref{eigenvalli}) suffices to explain the entries
in the eighth column of table \ref{cohomotab}. Indeed we can write:
\begin{equation}
\mbox{dim} \, \mbox{Im} \, \partial_3 = 20 \, - \, 2 \times \left( \mbox{$\#$ of null
skew eigenvalues of $\mathcal{T}$}\right)
\label{xitrii}
\end{equation}
This concludes my discussion of the cohomology groups
\subsection{The orthogonal basis of $p$ cochains in one example}
\label{orthogosec}
As a further illustration of the general set up I choose the case of
the SS algebra with generic non vanishing eigenvalues $m_1 \ne m_2
\ne m_3 \ne 0$ and for such an algebra I perform the explicit
construction of the orthogonal basis introduced in table
\ref{unciricamo}. In this case table \ref{cohomotab} takes the
following explicit form:
\begin{equation}
  \begin{array}{||c||c|c|c||}
  \hline
  \hline
    n & h_n & \wp_n & r_n \\
    \hline
    \hline
    7 & 1 & 0 & 0 \\
    \hline
    6 & 1 & 6 & 0 \\
    \hline
    5 & 3 & 12 & 6 \\
    \hline
    4 & 3 & 20 & 12 \\
    \hline
    3 & 3 & 12 & 20 \\
    \hline
    2 & 3 & 6 & 12 \\
    \hline
    1 & 1 & 0 & 6 \\
    \hline
    0 & 1 & 0 & 0 \\
    \hline
  \end{array}
\label{tavoletta}
\end{equation}
Let me construct one by one all the elements of the basis. I start
from $\Gamma^{[7]}$. This is just the volume form:
\begin{equation}
  \Gamma^{[7]} = \mathrm{Vol} = \frac{1}{7!} \epsilon_{I_1I_2 \dots
  I_7} e^{I_1} \wedge e^{I_2} \wedge \dots \wedge e^{I_7}
\label{gammasette}
\end{equation}
Its dual is the number $1$:
\begin{equation}
  \Gamma^{[0]} = 1
\label{tuttouno}
\end{equation}
which indeed fulfills the relation:
\begin{equation}
  \Gamma^{[7]} \, \wedge \, \Gamma^{[0]} = \mathrm{Vol}
\label{lapalissa}
\end{equation}
Next I turn to the pair $ \Gamma^{[6]} $ and $\Gamma^{[1]}$. Here we
have $h_1=h_6=1$ and it suffices to pose:
\begin{equation}
  \Gamma^{[1]} = e^0 \quad ; \quad \Gamma^{[6]} = \frac{1}{6!} \,
  \epsilon_{i_1 \, \dots \, i_6} \, e^{i_1} \wedge \dots \wedge
  e^{i_6} \quad \Rightarrow \quad \Gamma^{[1]} \, \wedge \,
  \Gamma^{[6]} \, = \, \mathrm{Vol}
\label{gamma1gamma6}
\end{equation}
To complete the analysis in dimensions $1$ and $6$, I have to
construct the pairs $\Xi^{[1]}_{\bar{\alpha}}$ and
$\Xi^{[5]}_{\bar{\beta}}$,
where ${\bar{\alpha}},{\bar{\beta}}=1,\dots,6$. To this effect
let me consider the symmetric, non degenerate $ 6 \times 6$ matrix:
\begin{equation}
  \mathcal{M} = 5! \, T^2
\label{matroxxa}
\end{equation}
and let $\overrightarrow{w}_{\bar{\alpha}}$ be an orthonormal system
of $6$ vectors with respect to the scalar product defined by
$\mathcal{M}$:
\begin{equation}
  \overrightarrow{w}_{\bar{\alpha}} \cdot
  \overrightarrow{w}_{\bar{\beta}} \equiv {w}_{\bar{\alpha}p} \,
  \mathcal{M}^{p}_{\phantom{p} r } \, {w}_{\bar{\beta}}^r \, = \, \delta_{ \bar{\alpha}\bar{\beta}}
\label{olidata}
\end{equation}
In terms of these vectors I set:
\begin{eqnarray}
\Xi^{[1]}_{\bar{\alpha}} & = & w_{\bar{\alpha}|p} \, T^{p}_{\phantom{p}r} \, e^r \nonumber\\
\Xi^{[5]}_{\bar{\beta}} & = & w_{\bar{\alpha}}^i \, \epsilon_{ij_1 \dots j_5} \,
e^{j_1} \, \wedge \dots \wedge \, e^{j_5}
\label{xi1xi5}
\end{eqnarray}
and I immediately obtain:
\begin{equation}
  \partial \, \Xi^{[5]}_{\bar{\beta}}  = 5\, e^0 \wedge  w_{\bar{\beta}}^i \, \epsilon_{ij_1 \dots j_5} \,
e^{j_1} \, \wedge \dots \wedge \, e^{j_5}
\label{partinico}
\end{equation}
Then by means of straightforward algebraic manipulations, you can
verify that:
\begin{equation}
  \Xi^{[1]}_{\bar{\alpha}} \, \wedge \, \partial \Xi^{[5]}_{\bar{\beta}}
  = \overrightarrow{w}_{\bar{\alpha}} \cdot
  \overrightarrow{w}_{\bar{\beta}} \, \mathrm{Vol}
\label{orthogone15}
\end{equation}
as requested.
\par
The next step is the construction of the pair $\Gamma^{[2]}_a$ ,
$\Gamma^{[5]}_b$. To this effect let $\sigma_{a}$ be a basis of three $6
\times 6$ matrices in the CSA of $\mathrm{SO(6)}$, commuting, by
definition, with $T$ and normalized in the following way:
\begin{equation}
  \mbox{Tr} \left( \sigma_a \, \sigma_b \right)  = - \frac{1}{4!2!}
  \, \delta_{ab}
\label{normalna}
\end{equation}
Then I pose:
\begin{eqnarray}
\Gamma^{[2]}_a & = & \sigma_{a|j_1j_2} \, e^{j_1} \wedge e^{j_2} \nonumber\\
\Gamma^{[5]}_b & = & e^0 \, \wedge \, \epsilon_{i_1\dots i_4 uv} \,
\sigma_b^{uv} \, e^{i_1} \wedge \dots \wedge \, e^{i_4}
\label{tentenna}
\end{eqnarray}
and we have both $\partial \Gamma^{[2]}_a = 0$ as we already know, and
$\partial \Gamma^{[5]}_b = 0$ for the simple reason that there is
already one $e^0$ and the boundary operator introduces a second.
Furthermore one easily verifies that:
\begin{equation}
  \Gamma^{[2]}_a \, \wedge \, \Gamma^{[5]}_b = -4!2! \, \times \, \mbox{Tr}\left(
  \sigma_a \, \sigma_b \right)  \, \mathrm{Vol}
\label{gamma2gamma5}
\end{equation}
Let me now turn my attention to the construction of the pair $\Xi^{[2]}_{\bar{a}}$ and
$\Xi^{[4]}_{\bar{b}}$. For this purpose let me recall that the
remaining $12$ generators of $\mathrm{SO(6)}$, which are not in the
CSA can be organized into a set of matrices $E_{\bar{a}}$ fulfilling
the relations:
\begin{eqnarray}
\left[ T, E_{\bar{a}}\right ]  & =  & K_{\bar{a} \bar{b}} \, E_{\bar{b}} \nonumber\\
\mbox{Tr} \left( E_{\bar{a}}E_{\bar{b}}\right)  & = &
\delta_{\bar{a}\bar{b}}
\label{genico}
\end{eqnarray}
where $K_{\bar{a} \bar{b}}$ is an antisymmetric $12 \times 12$
matrix. Indeed $E_{\bar{a}}$  are constructed with the step operators associated with roots
 of the $\mathrm{SO(6)}$ Lie algebra and $T$ is an element of the CSA. Then I can just pose:
\begin{eqnarray}
\Xi^{[2]}_{\bar{a}} & = & E_{\bar{a}|i_1i_2} \, e^{i_1} \wedge e^{i_2} \nonumber\\
\Xi^{[4]}_{\bar{b}} & = &  \frac{1}{4!2!} \,\left(  K^{-1}\right)
_{\bar{b}\bar{c}} \, \epsilon_{j_1 \dots j_4 uv} E_{\bar{c}}^{uv} \,
e^{j_1} \wedge \dots \wedge e^{j_4}
\label{xi2xi4}
\end{eqnarray}
Applying the boundary operator I find:
\begin{equation}
  \partial \, \Xi^{[4]}_{\bar{b}} \,  =\, \frac{1}{3!2!} \,e^0 \wedge \, T^{j_1}_{\phantom{j_1}r} \, \left(  K^{-1}\right)
_{\bar{b}\bar{c}} \, \epsilon_{j_1 \dots j_4 uv} E_{\bar{c}}^{uv} \,
\e^r \wedge \, e^{j_2} \wedge \dots \wedge e^{j_4}
\label{dxi4}
\end{equation}
and by direct computation I obtain:
\begin{eqnarray}
\Xi^{[2]}_{\bar{a}}\, \wedge \, \partial \, \Xi^{[4]}_{\bar{b}} & = &  \left(  K^{-1}\right)
_{\bar{b}\bar{c}} \, \mbox{Tr} \left( \left[ E_{\bar{a}} \, , \,  T\right] E_{\bar{c}}\right)
\, \times \,\mathrm{Vol} = \delta_{\bar{a}\bar{b}} \, \times \, \mathrm{Vol}
\label{ortox2xi4}
\end{eqnarray}
as requested.
\par
It remains, to be studied, the pair of sectors in grade $3$ and $4$. The
basis for the space $\mathbf{\Xi}^{[4]}$ has already been constructed
in eq.(\ref{xi2xi4}). As for the basis of the dual spaces
$\mathbf{\Gamma}^{[3]}$ and $\mathbf{\Gamma}^{[4]}$ it suffices to
pose:
\begin{eqnarray}
\Gamma_x^{[4]} & = & \epsilon_{i_1 \dots i_4 uv} \, \sigma^{uv}_x \, e^{i_1} \wedge \dots \wedge e^{i_4} \nonumber\\
\Gamma_y^{[3]} & = & e^0 \, \wedge \, \sigma_{y|uv} \, e^u \wedge e^v
\label{lubiapitni}
\end{eqnarray}
where $\sigma_x$ are the same $6 \times 6 $ matrices in the CSA of
$\mathrm{SO(6)}$ introduced in equation (\ref{normalna}). You can
immediately verify that in this way the forms are already correctly
normalized:
\begin{equation}
  \Gamma_y^{[3]} \, \wedge \, \Gamma_x^{[4]} \, = \, \delta_{xy} \,
  \times \, \mathrm{Vol}
\label{bestiona}
\end{equation}
Finally I have to define the basis for the space $\mathbf{\Xi}^{[3]}$
its conjugate being in this case $\partial \mathbf{\Xi}^{[3]}$. To
this effect let me recall the $20 \times 20$ matrix $\mathcal{T}$
introduced in eq. (\ref{Taumatrone}). Formally it was defined in the
following way. Let $U_{i_1i_2i_3|\bar{x} }$ be the lexicographic basis of rank
three antisymmetric tensors with $\bar{x} = 1,\dots ,20$  normalized as follows:
\begin{equation}
U_{i_1i_2i_3|\bar{x} } U^{j_1j_2j_3}_{\bar{y} } \, \epsilon_{i_1i_2i_3j_1j_2j_3} \, = \,
\Omega_{\bar{x}\bar{y}}
\label{ordinamento}
\end{equation}
where $\Omega_{\bar{x}\bar{y}}$ is some $20 \times 20$ non degenerate antisymmetric matrix.
Then I have defined:
\begin{equation}
  T^{r}_{\phantom{r}[i} \,U_{jk]r|\bar{x} } =
  \mathcal{T}_{\bar{x}\bar{y} } \, U_{ijk |\bar{y}}
\label{taino}
\end{equation}
and I have obtained the explicit result (\ref{Taumatrone}). Let me now
set:
\begin{equation}
  \Xi^{[3]}_{\bar{x}} = U_{i_1i_2i_3|\bar{x} } \, e^{i_1} \, \wedge
  \, e^{i_2} \, \wedge \,  e^{i_3}
\label{Xi3}
\end{equation}
A straightforward calculation immediately shows that:
\begin{equation}
 \partial \,  \Xi^{[3]}_{\bar{x}} \, \wedge \, \Xi^{[3]}_{\bar{y}} =
 \left\{ \Omega \, , \, \mathcal{T}\right\} _{\bar{x}\bar{y}} \,
 \times \, \mathrm{Vol}
\label{megabritto}
\end{equation}
The matrix $\left\{ \Omega \, , \, \mathcal{T}\right\}$ is symmetric,
being the anticommutator of two antisymmetric matrices and non
degenerate. By a suitable change of basis we can always reduce it to
a be a delta.
\par
This concludes my discussion and illustration of
Chevalley cohomology with a particular attention to the aspects
relevant for M--theory compactification on twisted tori.
\par
I next turn to recall the very concept and the structural theorems
relative to free differential algebras.
\section{General Structure of FDA.s and Sullivan's theorems}
\label{sullivano}
As I have already recalled in the introduction Free Differential Algebras (FDA) are a natural categorical extension of the
notion of Lie algebra and constitute the natural mathematical
environment for the description of the algebraic structure of higher
dimensional supergravity theory, hence also of string theory. The
reason is the ubiquitous presence in the spectrum of
string/supergravity theory of antisymmetric gauge fields ($p$--forms)
of rank greater than one.
\par
FDA.s were independently discovered in Mathematics by Sullivan \cite{sullivan}
and in Physics by the present author in collaboration with R. D'Auria
\cite{fredauria}. The original name given to this algebraic structure
by D'Auria and me was that of \textit{Cartan Integrable Systems}.
Later, recognizing the conceptual identity of our supersymmetric construction with the pure
bosonic constructions considered by Sullivan, we also turned to its naming FDA which has by now become
generally accepted. In this section my purpose is that of recalling
the definition of FDA.s and two structural theorems by Sullivan which
show how all possible FDA.s are, in a sense to be described,
\textit{cohomological extensions} of normal Lie algebras or
superalgebras.
\par
Another question which is of utmost relevance in all physical
applications is that of \textit{gauging} of FDA.s. Just in the same
way as physics gauges standard Lie algebras by means of Yang
Mills theory through the notion of gauge connections and curvatures one
expects to gauge FDA.s by introducing  their curvatures. A
surprising feature of the FDA setup which was noticed and explained
by me in a paper of 1985 \cite{comments} is that differently from Lie
algebras the algebraic structure of FDA already encompasses both the
notion of connection and the notion of curvature and there is a
well defined mathematical way of separating the two which relies on
the two structural theorems by Sullivan. In this section I ultra shortly summarize
this essential mathematical lore as a preparation for my analysis of
twisted tori compactifications. It is indeed my goal to show how the
recent findings in this realm fit and are properly interpreted within
the cohomological set-up of Chevalley cohomology and the
cohomological classification scheme of FDA.s Adopting this language
and its appropriate technical tools will also allow me to correct
some imprecise statements appeared in the literature and provide a
general scheme for a systematic analysis of all cases of interest.
\par
\paragraph{Definition of FDA} The starting point for FDA.s is the
generalization of Maurer Cartan equations. As we already emphasized
in section \ref{chevalley} a standard Lie algebra is defined by its
structure constants which can be alternatively introduced either
through the commutators of the generators as in eq.(\ref{commrel}) or
through the Maurer Cartan equations obeyed by the dual $1$--forms as
in eq.(\ref{MC1}). The relation between the two descriptions is
provided by the duality relation in eq. (\ref{dualityrele}). Adopting
the Maurer Cartan viewpoint FDA.s can now be defined as follows.
Consider a formal set of exterior forms $\left\{ \theta^{A(p)}\right\}
$ labelled by the index $A$ and by the degree $p$ which may be
different for different values of $A$. Given this set we can write a
set of generalized \textit{Maurer Cartan equations} of the following type:
\begin{equation}
  d\theta^{A(p)} \, + \, \sum_{n=1}^{N} \,
  C^{A(p)}_{\phantom{{A(p)}}B_1(p_1)\dots B_n(p_n)} \,
  \theta^{B_1(p_1)} \, \wedge \, \dots \, \wedge \theta^{B_n(p_n)}\, = \,0
\label{detheta}
\end{equation}
where $C^{A(p)}_{\phantom{{A(p)}}B_1(p_1)\dots B_n(p_n)}$ are generalized structure constants with the
  same symmetry as induced by permuting the $\theta$.s in the wedge
  product. They can be non--zero only if:
\begin{equation}
  p+1= \sum_{i=1}^n \, p_i
\label{p+1}
\end{equation}
Equations (\ref{detheta}) are self-consistent and define an FDA if
and only if $dd\theta^{A(p)}=0$ upon substitution of (\ref{detheta})
into its own derivative. This procedure yields the generalized Jacobi
identities of FDA.s.\footnote{For a review of FDA theory see
\cite{castdauriafre}}
\paragraph{Classification of FDA and the analogue of Levi theorem:  minimal versus contractible algebras} A
basic theorem of Lie algebra theory states that the most general Lie
algebra $\mathcal{A}$ is the semidirect product of a semisimple Lie
algebra $\mathcal{L}$ called the Levi subalgebra with
$\mbox{Rad}(\mathcal{A})$, namely with the radical of $\mathcal{A}$. By
definition this latter is the maximal solvable ideal of $\mathcal{A}$.
Sullivan \cite{sullivan} has provided an analogous structural theorem
for FDA.s. To this effect one needs the notions of \textit{minimal
FDA} and \textit{contractible FDA}. A minimal FDA is one for which:
\begin{equation}
  C^{A(p)}_{\phantom{A(p)}B(p+1)} \, = \, 0
\label{minimalconst}
\end{equation}
This excludes the case where a $(p+1)$--form appears in the
generalized Maurer Cartan equations as a contribution to the
derivative of a $p$--form. In a minimal algebra all non differential
terms are products of at least two elements of the algebra, so that
all forms appearing in the expansion of $d\theta^{A(p)}$ have at most
degree $p$, the degree $p+1$ being ruled out.
\par
On the other hand a \textit{contractible FDA} is one where the only
form appearing in the expansion of $d\theta^{A(p)}$ has degree $p+1$,
namely:
\begin{equation}
  d\theta^{A(p)} = \theta^{A(p+1)} \quad \Rightarrow \quad
  d\theta^{A(p+1)} = 0
\label{contraFDA}
\end{equation}
A contractible algebra has a trivial structure. The basis $\left\{
\theta^{A(p)}\right\}$ can be subdivided in two subsets $\left\{ \Lambda^{A(p)}\right\} $
and $\left\{ \Omega^{B(p+1)}\right\} $ where $A$ spans a subset of the values taken by
$B$, so that:
\begin{equation}
  d\Omega^{B(p+1)} = 0
\label{forallB}
\end{equation}
for all values of $B$ and
\begin{equation}
   d\Lambda^{A(p)}= \Omega^{A(p+1)}
\label{unguedA=B}
\end{equation}
Denoting by $\mathcal{M}^k$ the vector space generated by all forms
of degree $p\le k$ and $C^k$ the vector space of forms of degree $k$,
a minimal algebra is shortly defined by the property:
\begin{equation}
  d\mathcal{M}^k \, \subset \, \mathcal{M}^k \wedge \mathcal{M}^k
\label{lottominimo}
\end{equation}
while a contractible algebra is defined by the property
\begin{equation}
  d C^k \, \subset \, C^{k+1}
\label{lottocontrat}
\end{equation}
In analogy to Levi's theorem, the first theorem by Sullivan states
that: \textit{The most general FDA is the semidirect sum of a
contractible algebra with a minimal algebra}
\par
\paragraph{Sullivan's first theorem and the gauging of FDA.s}
Twenty years ago in \cite{comments} I observed that the above
mathematical theorem has a deep physical meaning relative to the
gauging of algebras. Indeed I proposed the following identifications:
\begin{enumerate}
  \item The \textit{contractible generators} $\Omega^{A(p+1)}+\dots$ of any given FDA $\mathbb{A}$ are
  to be physically identified with the \textit{curvatures}
  \item The Maurer Cartan equations that begin with
  $d\Omega^{A(p+1)}$ are \textit{the Bianchi identities}.
  \item The algebra which is gauged is the \textit{minimal
  subalgebra} $\mathbb{M} \subset \mathbb{A} $.
  \item{The Maurer Cartan equations of the minimal subalgebra
  $\mathbb{M}$ are consistently obtained by those of $\mathbb{A}$ by setting
  all contractible generators to zero.}
\end{enumerate}
\par
\paragraph{Sullivan's second structural theorem and Chevalley cohomology}
The second structural theorem proved by Sullivan\footnote{For detailed explanations on this see
again, apart from the original article \cite{sullivan} the book \cite{castdauriafre}} deals with the
structure of minimal algebras and it is constructive. Indeed it
states that the most general minimal FDA $\mathbb{M}$ necessarily
contains an ordinary Lie subalgebra $\mathbb{G} \subset \mathbb{M}$
whose associated $1$--form generators we can call $e^I$, as in
equation (\ref{MC1}). Additional $p$--form generators $A^{[p]}$ of
$\mathbb{M}$ are necessarily, according to Sullivan's theorem, in one--to--one correspondence with
Chevalley $p+1$ cohomology classes $\Gamma^{[p+1]}\left (e\right)$ of $\mathbb{G}\subset \mathbb{M}$.
Indeed, given such a class, which is a polynomial in the $e^I$
generators, we can consistently write the new higher degree Maurer
Cartan equation:
\begin{equation}
  \partial \, A^{[p]} + \Gamma^{[p+1]}(e) = 0
\label{oppato}
\end{equation}
where $A^{[p]}$ is a new object that cannot be written as a
polynomial in the old objects $e^I$. Considering now the FDA generated
by the inclusion of the available $A^{[p]}$, one can inspect its
Chevalley cohomology: the cochains are the polynomials in the extended set of
forms $\left \{ A,e^I\right \}$ and the boundary operator is defined
by the enlarged set of Maurer Cartan equations. If there are new
cohomology classes $\Gamma^{[p+1]}\left( e,A \right)$, then one can further
extend the FDA by including new $p$--generators $B^{[p]}$ obeying the
Maurer Cartan equation:
\begin{equation}
  \partial \, B^{[p]} + \Gamma^{[p+1]}\left (e,A\right) = 0
\label{oppato2}
\end{equation}
The iterative procedure can now be continued by inspecting the
cohomology classes of type $\Gamma^{[p+1]}\left (e,A,B\right)$ which
lead to new generators $C^{[p]}$ and so on. Sullivan's theorem states that
those constructed in this way are, up to isomorphisms, the most
general minimal FDA.s.
\par
To be precise, this is not the whole story. There is actually one
 generalization that should be taken into account. Instead of
 \textit{absolute Chevalley cohomology} one can rather consider
 \textit{relative Chevalley cohomology}. This means that rather then
being $\mathbb{G}$- singlets, the Chevalley $p$-cochains can be assigned to
some linear representation of the Lie algebra $\mathbb{G}$. In this
case eq.(\ref{pcochain}) is replaced by:
\begin{equation}
  \Omega^{\alpha[p]} = \Omega_{I_1\dots I_p}^\alpha \, e^{I_1} \, \wedge \, \dots
  \wedge \, e^{I_p}
\label{pcochain2}
\end{equation}
where the index $\alpha$ runs in some representation $D$:
\begin{equation}
  D \quad : \quad T_I \, \rightarrow \,
  \left[D\left(T_I\right)\right]^\alpha_{\phantom{\alpha}\beta}
\label{onice}
\end{equation}
 and the boundary operator is now the covariant $\nabla$:
\begin{eqnarray}
  \nabla \, \Omega^{\alpha[p]} & \equiv & \partial \Omega^{\alpha[p]}
  \, +\, e^I \, \wedge \,
  \left[D\left(T_I\right)\right]^\alpha_{\phantom{\alpha}\beta} \,
  \Omega^{\beta[p]}
\label{dcov}
\end{eqnarray}
Since  $\nabla^2=0$, we can repeat all previously explained steps and compute cohomology groups.
Each non trivial cohomology class $\Gamma^{\alpha[p+1]}(e)$ leads to
new $p$--form generators $A^{\alpha[p]}$ which are assigned to the
same $\mathbb{G}$--representation as $\Gamma^{\alpha[p+1]}(e)$. All
successive steps go through in the same way as before and Sullivan's
theorem actually states that all minimal FDA.s are obtained in this
way for suitable choices of the representation $D$, in particular the
singlet.
\subsection{Non trivial FDA extensions of the SS algebras and anticipations on twisted tori
compactifications of M--theory}
\label{minimSS}
In sections \ref{Sscohomo} and \ref{orthogosec} I presented a
detailed study of the cohomology groups for the Scherk--Schwarz
algebras (\ref{Gexemplo}), which are of interest in M--theory
compactifications on twisted tori. In order to illustrate the bearing
of Sullivan's structural theorems, let me describe the minimal FDA.s that
could be constructed starting from such algebras $\mathbb{G}$. For
instance, given the cohomology groups, one could extend the original
algebra with $h_2(\mathbb{G})$ new $1$--form generators $A_{a}^{[1]}$, by
writing:
\begin{equation}
  dA_{a}^{[1]} + \Gamma^{[2]}_a=0 \quad ; \quad a=1,\dots , h_2(\mathbb{G})
\label{exemplovero}
\end{equation}
Similarly one could extend the FDA with
$h_3(\mathbb{G})$ two-forms $A_{x}^{[2]}$  by writing:
\begin{equation}
   dA_{x}^{[2]} + \Gamma^{[3]}_x=0 \quad ; \quad x=1,\dots , h_3(\mathbb{G})
\label{exemplovero2}
\end{equation}
In components the above non--trivial minimal algebra would lead to
the following curvature definitions, upon extension with the contractible
generators according to first Sullivan's theorem:
\begin{eqnarray}
G^I & \equiv & dW^I + \ft 12 \tau^{I}_{\phantom{I}JK} \, W^J \, \wedge \, e^K \nonumber\\
F^{[2]}_a & \equiv & dA_{a}^{[1]} + \Gamma^{[2]}_{a|IJ} \, W^I \, \wedge \,
W^J \nonumber\\
F^{[3]}_x & \equiv & dA_{x}^{[2]} + \Gamma^{[3]}_{x|IJK} \, W^I \, \wedge
\, W^J \, \wedge \, W^K
\label{WF}
\end{eqnarray}
Note that in the above algebra the index carried by the new $p$--form
generators is lower just as in the cohomology class that generates it.
I stress this because it turns out that in the FDA.s algebras generated
by twisted tori compactifications this will not be the case and those FDA.s will in general be different
from the above. Sullivan's structural theorem creates an association between
cohomology classes and  new forms $p$ generators of the FDA which goes
as follows:
\begin{equation}
  \forall \, \Gamma^{(p+1)}_x \, \in \, \mathrm{H}^{(p+1)}\left( \mathbb{G}\right)
  \, \Rightarrow \, \exists \, \mbox{new generator} \, A^{[p]}_x \, \mbox{of
  degree} \, p
\label{Sullivanpredi}
\end{equation}
Also in M--theory compactifications there is an association
between cohomology classes and new generators of the FDA but, as we will see, it
rather  as follows:
\begin{equation}
  \forall \, \Gamma^{(p)}_x \, \in \, \mathrm{H}^{(p)}\left( \mathbb{G}\right)
  \, \Rightarrow \, \exists \, \mbox{new generator} \, \Sigma_{[3-p]}^x \, \mbox{of
  degree} \, 3-p
\label{Mpredi}
\end{equation}
Eq. (\ref{Mpredi}) takes origin by the fact that  $\Sigma_{[3-p]}^x \,
\wedge \, \Gamma^{(p)}_x$ will a contribution to the development of the
three--form potential of M--theory. For this reason the index carried
by the new generator is not in the same lower position as in the cohomology
class rather it is in the upper position, \textit{i.e.} transforms
controvariantly with respect to any symmetry that acts on Chevalley
cocycles.
\par
On the other hand Sullivan's structural theorem holds true for all
FDA.s, also those  emerging from twisted tori compactifications.
Hence the new generators $\Sigma_{[3-p]}^x$ associated with
$p$--cohomology classes can be non trivial, namely not contractible
ones, if and only if they are suitable paired with $(4-p)$ Chevalley
cocycles $\Gamma^{(4-p)}_x$ which, from M--theory reduction, should
automatically appear in the appropriate position as second member of
the equation. In other words, from such reductions we obtain a non trivial FDA, if and only if:
\begin{eqnarray}
\Gamma^{(p)}_x \, \in \, \mathrm{H}^{(p)}\left( \mathbb{G}\right)
  \, \Rightarrow \,  \Sigma_{[3-p]}^x \, \nonumber\\
  d\Sigma_{[3-p]}^x & = & \Gamma^{(4-p)}_x \, \in \, \mathrm{H}^{(p)}\left( \mathbb{G}\right)
\label{nontrivialone}
\end{eqnarray}
It will turn out that, generically, the situation
(\ref{nontrivialone}) is not realized. There are new generators
associated with cohomology classes but they are in general contractible
and the minimal FDA is nothing else but the original algebra $\mathbb{G}$.
I postpone an enlarged discussion of this point to the end of section
\ref{introcurve} after deriving the details of M--theory reduction.
\section{The super FDA of M theory and its cohomological structure}
\label{superFDA}
Sullivan's theorems have been introduced and proved for Lie algebras
and their corresponding FDA extensions but they hold true, with obvious
modifications, also for superalgebras ${\mathbb{G}}_s$ and for their
FDA extensions. Actually, in view of superstring and supergravity, it
is precisely in the supersymmetric context that FDA.s have found
their most relevant applications. In this section, as an illustration
of the general set up and also as an introduction to my specific
interest, which is M--theory compactifications, I present the
structure of the M--theory FDA, by recalling the results of \cite{fredauria}
and \cite{comments}. Within this context I will also be able to
illustrate the bearing of a quite relevant question: \textit{is an
FDA  $\mathbb{M}$ always equivalent to a normal Lie algebra $\widehat{\mathbb{G}}
\supset \mathbb{G}$ larger than the Lie algebra of which $\mathbb{M}$ is a
cohomological extension?}. How to mathematically formulate and answer
such a question I will show below by recalling results of
\cite{fredauria} and also more recent literature
\cite{spagnoli1,spagnoli2,nastase}.
\par
Let me begin by  writing the complete set of curvatures, plus their Bianchi identities. This
will define the complete FDA:
\begin{equation}
  \mathbb{A}=\mathbb{M}\biguplus\mathbb{C}
\label{completealg}
\end{equation}
The curvatures being the contractible generators $\mathbb{C}$. By setting them to zero we  retrieve,
according to Sullivan's first theorem, the minimal
algebra $\mathbb{M}$. This latter, according instead to Sullivan's second
theorem, has to be explained in terms of cohomology of the
 normal subalgebra $\mathbb{G} \subset \mathbb{M}$, spanned by the
 $1$--forms. In this case  $\mathbb{G}$ is just
  the $D=11$ superalgebra spanned by the following $1$--forms:
\begin{enumerate}
  \item the vielbein $V^a$
  \item the spin connection $\omega^{ab}$
  \item the gravitino $\psi$
\end{enumerate}
The higher degree generators of the minimal FDA $\mathbb{M}$ are:
\begin{enumerate}
  \item the bosonic $3$--form $\mathbf{A^{[3]}}$
  \item the bosonic $6$-form $\mathbf{A^{[6]}}$.
\end{enumerate}
The complete set of curvatures is given below (\cite{fredauria,comments}):
\begin{eqnarray}
T^{a} & = & \mathcal{D}V^a - {\rm i} \ft 12 \, \overline{\psi} \, \wedge \, \Gamma^a \, \psi \nonumber\\
R^{ab} & = & d\omega^{ab} - \omega^{ac} \, \wedge \, \omega^{cb}
\nonumber\\
\rho & = & \mathcal{D}\psi \equiv d \psi - \ft 14 \, \omega^{ab} \, \wedge \, \Gamma_{ab} \, \psi\nonumber\\
\mathbf{F^{[4]}} & = & d\mathbf{A^{[3]}} - \ft 12\, \overline{\psi} \, \wedge \, \Gamma_{ab} \, \psi \,
\wedge \, V^a \wedge V^b \nonumber\\
\mathbf{F^{[7]}} & = & d\mathbf{A^{[6]}} -15 \, \mathbf{F^{[4]}} \, \wedge \,  \mathbf{A^{[3]}} - \ft {15}{2} \,
\, V^{a}\wedge V^{b} \, \wedge \, {\bar \psi}
\wedge \, \Gamma_{ab} \, \psi
\, \wedge \, \mathbf{A^{[3]}} \nonumber\\
\null & \null & - {\rm i}\, \ft {1}{2} \, \overline{\psi} \, \wedge \, \Gamma_{a_1 \dots a_5} \, \psi \,
\wedge \, V^{a_1} \wedge \dots \wedge V^{a_5}
\label{FDAcompleta}
\end{eqnarray}
From their very definition, by taking a further exterior derivative
one obtains the Bianchi identities, which for brevity I do not
explicitly write (see \cite{comments}). The dynamical theory is
defined, according to a general constructive scheme of supersymmetric theories, by the principle
of rheonomy (see \cite{castdauriafre} ) implemented
into Bianchi identities.
Indeed there is a unique rheonomic parametrization of the curvatures (\ref{FDAcompleta}) which solves the
Bianchi identities and it  is the following one:
\begin{eqnarray}
T^a & = & 0 \nonumber\\
\mathbf{F^{[4]}} & = & F_{a_1\dots a_4} \, V^{a_1} \, \wedge \dots \wedge \, V^{a_4} \nonumber\\
\mathbf{F^{[7]}} & = & \ft {1}{84} F^{a_1\dots a_4} \, V^{b_1} \, \wedge \dots \wedge \,
V^{b_7} \, \epsilon_{a_1 \dots a_4 b_1 \dots b_7} \nonumber\\
\rho & = & \rho_{a_1a_2} \,V^{a_1} \, \wedge \, V^{a_2} - {\rm i} \ft 12 \,
\left(\Gamma^{a_1a_2 a_3} \psi \, \wedge \, V^{a_4} + \ft 1 8
\Gamma^{a_1\dots a_4 m}\, \psi \, \wedge \, V^m
\right) \, F^{a_1 \dots a_4} \nonumber\\
R^{ab} & = & R^{ab}_{\phantom{ab}cd} \, V^c \, \wedge \, V^d
+ {\rm i} \, \rho_{mn} \, \left( \ft 12 \Gamma^{abmn} - \ft 2 9 \Gamma^{mn[a}\, \delta^{b]c} + 2 \,
\Gamma^{ab[m} \, \delta^{n]c}\right) \, \psi \wedge V^c\nonumber\\
 & &+\overline{\psi} \wedge \, \Gamma^{mn} \, \psi \, F^{mnab} + \ft 1{24} \overline{\psi} \wedge \,
 \Gamma^{abc_1 \dots c_4} \, \psi \, F^{c_1 \dots c_4}
\label{rheoFDA}
\end{eqnarray}
The expressions (\ref{rheoFDA}) satisfy the Bianchi.s provided the space--time components
of the curvatures satisfy the following constraints
\begin{eqnarray}
0 & = & \mathcal{D}_m F^{mc_1 c_2 c_3} \, + \, \ft 1{96} \, \epsilon^{c_1c_2c_3 a_1 a_8} \, F_{a_1 \dots a_4}
\, F_{a_5 \dots a_8}  \nonumber\\
0 & = & \Gamma^{abc} \, \rho_{bc} \nonumber\\
R^{am}_{\phantom{bm}cm} & = & 6 \, F^{ac_1c_2c_3} \,F^{bc_1c_2c_3} -
\, \ft 12 \, \delta^a_b \, F^{c_1 \dots c_4} \,F^{c_1 \dots c_4}
\label{fieldeque}
\end{eqnarray}
which are the  space--time field equations.
\subsection{The minimal FDA and cohomology}
Setting $T^a=R^{ab}=\rho=\mathbf{F^{[4]}}=\mathbf{F^{[7]}}=0$ in eq.s
(\ref{FDAcompleta}) we obtain the Maurer Cartan equations of the
minimal algebra $\mathbb{M}$. In particular we have:
\begin{eqnarray}
d\mathbf{A^{[3]}}&=& \mathbf{\Gamma^{[4]}}\left( V,\psi\right) \, \equiv \, \ft 12\,
\overline{\psi} \, \wedge \, \Gamma_{ab} \, \psi \,
\wedge \, V^a \wedge V^b \nonumber\\
d\mathbf{A^{[6]}} & = & \mathbf{\Gamma^{[7]}}\left( V,\psi, \mathbf{A^{[3]}}\right) \nonumber\\
& \equiv &
\ft {15}{2} \, \, V^{a}\wedge V^{b} \,
\wedge \,\overline{\psi} \, \wedge \, \Gamma_{ab} \, \psi
\, \wedge \, \mathbf{A^{[3]}}  + {\rm i}\, \ft {1}{2} \, \overline{\psi} \, \wedge \, \Gamma_{a_1 \dots a_5} \, \psi \,
\wedge \, V^{a_1} \wedge \dots \wedge V^{a_5}
\label{cohomolosuperFDA}
\end{eqnarray}
The reason why the three--form generator $\mathbf{A^{[3]}}$ does exist
and also why the six--form generator $\mathbf{A^{[6]}}$ can be
included is, in this set up, a direct consequence of the cohomology
of the super Poncar\'e algebra in $D=11$, via Sullivan's second
theorem. Indeed the $4$--form $\mathbf{\Gamma^{[4]}}\left(
V,\psi\right)$ defined in the first line of
eq.(\ref{cohomolosuperFDA}) is a cohomology class of the super
Poincar\'e Lie algebra whose Maurer Cartan equations are the first
three of eq.s (\ref{FDAcompleta}) upon setting $T^a=R^{ab}=\rho=0$.
We have:
\begin{equation}
  d \mathbf{\Gamma^{[4]}}\left( V,\psi\right) = 0
\label{coho4molMtheory}
\end{equation}
and there is no $\mathbf{\Phi^{[3]}}\left( V,\psi\right) $ such that
$\mathbf{\Gamma^{[4]}}\left( V,\psi\right)=d\mathbf{\Phi^{[3]}}\left(
V,\psi\right)$.
In this way we see that $M2$--branes and $M5$--branes, namely the
dynamical objects that make up M-theory and which respectively couple
to the forms $\mathbf{A^{[3]}}$ and $\mathbf{A^{[6]}}$ are just an
yield of Chevalley cohomology of the super Poincar\'e algebra.
\subsection{FDA equivalence with larger (super) Lie algebras}
I can now address the question I posed at the beginning of this
section. Are FDA.s eventually equivalent to normal (super) Lie
algebras? For minimal algebras the question can be nicely rephrased
in the following way: \textit{can a non trivial cohomology class of a Lie
algebra $\mathbb{G}$ be trivialized by immersing $\mathbb{G}$ into a larger
algebra $\widehat{\mathbb{G}}$?}  Indeed by adding new $1$--form generators $\phi^p$
which, together with the generators $e^I$ of $\mathbb{G}$ satisfy the
Maurer Cartan equations of the larger algebra $\widehat{\mathbb{G}}\, \supset \, \mathbb{G}$, it
may happen that we are able to construct a polynomial $\mathbf{\Phi}^{[p-1]}\left(
e,\phi\right) $ such that:
\begin{equation}
  d\mathbf{\Phi}^{[p-1]}\left(
e,\phi\right) = \mathbf{\Gamma}^{[p]}\left(
e \right)
\label{jorjesol}
\end{equation}
In this case the  generator $\mathbf{A^{[p-1]}}$ of the FDA associated with
the cohomology class $\mathbf{\Gamma}^{[p]}\left(
e \right)$ can be simply deleted by the list of independent
generators and simply identified with the polynomial $\mathbf{\Phi}^{[p-1]}\left(
e,\phi\right) $.
\par
In these terms the question was already posed twenty three years ago
by D'Auria and me in \cite{fredauria} obtaining a positive answer
\cite{fredauria} which has been recently revisited in
\cite{spagnoli1,spagnoli2}.
\par
The enlarged algebra $\widehat{\mathbb{G}}$ contains, besides the
generators of $\mathbb{G}$ a bosonic $1$--form $B^{a_1a_2}$ which is in the
rank two antisymmetric representation of the Lorentz group, a bosonic
$1$--form $B^{a_1a_2,\dots a_5}$ which is in the rank five
antisymmetric representation and finally a fermionic $1$--form $\eta$
which is in the spinor representation just as the generator $\psi$.
\par
The Maurer Cartan equations of $\widehat{\mathbb{G}}$ are:
\begin{eqnarray}
0 & = & R^{ab} \, \equiv \,  d\omega^{ab} \, - \, \omega^{ac} \, \wedge \, \omega^{cb}  \label{curvacurva}\\
0 & = & T^{a} \, \equiv \,  \mathcal{D} V^{a} - {\rm i}\ft 1 2 \, \overline{\psi} \, \wedge \, \Gamma^a \, \psi
\label{curvators1}\\
0 & = & T^{a_1a_2} \, \equiv \,  \mathcal{D} B^{a_1a_2}  - \ft 1 2 \, \overline{\psi}  \, \wedge \,
 \Gamma^{a_1a_2} \, \psi
\label{curvators2}\\
0 & = & T^{a_1\dots a_5} \, \equiv \,  \mathcal{D} B^{a_1 \dots a_5}
 - {\rm i}\ft 1 2 \, \overline{\psi}  \, \wedge \, \Gamma^{a_1 \dots a_5} \, \psi  \label{curvators5}\\
0 & = & \rho \, \equiv \,  \mathcal{D} \psi \, \equiv \, d\psi - \ft 1 4 \, \omega^{ab}  \, \wedge \, \Gamma^{ab} \, \psi
\label{curvarho}\\
0 & = & \sigma \, \equiv \,  \mathcal{D} \eta - {\rm i} \, \delta \, \Gamma^a \psi
 \wedge V^a -\gamma_1 \, \Gamma^{ab} \psi \wedge B^{ab} - \gamma_2 \,
\Gamma^{a_1 \dots a_5} \, \psi \wedge B^{a_1 \dots a_5}
\label{curvasigma}
\end{eqnarray}
These Maurer Cartan equations are consistent, namely closed provided
the following equation is satisfied by the coefficients:
\begin{equation}
  \delta + 10 \, \gamma_1 - 720 \, \gamma_2 = 0
\label{paraeq1}
\end{equation}
Using all the generators of $\widehat{\mathbb{G}}$ one can construct
a cubic polynomial
\begin{eqnarray}
 && \mathbf{\Phi^{[3]}}\left( V,\psi,B^{(2)},B^{(5)}\right) \, = \,
 \nonumber\\
 && \lambda B^{a_1a_2} \wedge V^{a_1} \wedge V^{a_2} +
  \alpha_1 \, B^{a_1a_2} \wedge B^{a2a3} \wedge B^{a_3a_1} \, + \,
  \alpha_2 \, B^{b_1a_1 \dots a_4 } \wedge B^{b_1b_2} \wedge
  B^{b_2a_1\dots a_4} \nonumber\\
  &&+\alpha _3 \, \epsilon ^{a_1 \dots a_5b_1 \dots b_4m} \, B^{a_1
  \dots a_5} \wedge B^{b_1 \dots b_5} \wedge V^m \nonumber\\
  && +\alpha _4 \epsilon ^{m_1\dots m_6 n_1 \dots n_5} \,
  B^{m_1m_2m_3 p_1p_2} \wedge B^{m_4m_5m_6 p_1p_2} \wedge B^{n_1\dots
  n_5} \nonumber\\
  && {\rm i} \, \beta_1 \,\overline{\psi} \wedge \Gamma^a \eta \wedge
  V^a + \beta_2 \, \overline{\psi}  \wedge \Gamma^{a_1a_2}
  \eta \wedge B^{a_1a_2} + {\rm i}\beta_3 \, \overline{\psi} \wedge \Gamma^{a_1
  \dots a_5} \eta \wedge B^{a_1 \dots a_5}
\label{A3form}
\end{eqnarray}
such that:
\begin{equation}
  d\mathbf{\Phi^{[3]}}\left( V,\psi,B^{(2)},B^{(5)}\right) =\mathbf{\Gamma^{[4]}}\left( V,\psi\right) \, \equiv \, \ft 12\,
\overline{\psi} \, \wedge \, \Gamma_{ab} \, \psi \,
\wedge \, V^a \wedge V^b
\label{trivializzo}
\end{equation}
The coefficients appearing in (\ref{A3form}) are completely fixed by eq.(\ref{trivializzo}) for any of the
$1$--parameter family of algebras described by
(\ref{curvacurva}-\ref{curvasigma}). Indeed the closure condition
(\ref{paraeq1}) is one equation on three parameters which are therefore reduced to two. One of them, say
$\gamma_1$ can be reabsorbed into the normalization of the extra
fermionic generator $\eta$, but the other remains essential and its
value selects one algebra within a family of non isomorphic ones.
Following \cite{spagnoli1} it is convenient to set:
\begin{equation}
  \delta = 2 \, \gamma_1 \, (s+1) \quad ; \quad \gamma_2 = 2 \,
  \gamma_1\, \left(\frac{s}{6!} + \frac{1}{5!}\right)
\label{esseparasol}
\end{equation}
and $s$ is the parameter which parametrizes the inequivalent algebras
$\widehat{\mathbb{G}}_s$. For each of them we have a solution of
eq.(\ref{trivializzo}) realized by
\begin{equation}
  \begin{array}{ccccccc}
    \alpha _1& = & {\frac{2\, \left( 3 + s \right) }{15\, s^2}} & ;
     & \alpha _4& = & {\frac{-{\left( 6 + s \right) }^2}{259200\, s^2}} \\
    \alpha _2& = & {\frac{-{\left( 6 + s \right)
}^2}{720\,s^2}} & ; & \alpha _3& = & {\frac{{\left( 6 + s \right)}^2}{432000\,s^2}}\\
   \lambda & = & {\frac{6 + 2\, s + s^2}{5\, s^2}} & ; &
   \beta _1& = & {\frac{3 - 2\, s}{10\, s^2\, {{\gamma}_1}}}\\
   \beta _2& = & {\frac{3 + s}{20\, s^2\, {{\gamma}_1}}} & ; &
   \beta _3 & = & {\frac{6 + s}{2400\, s^2\, {{\gamma}_1}}}\\
  \end{array}
\label{soluzi}
\end{equation}
In the original paper by D'Auria and myself \cite{fredauria} only two
of this infinite class of solutions were found, namely those
corresponding to the values:
\begin{equation}
  s = -1 \quad ; \quad s = \frac{3}{2}
\label{fredaucrit}
\end{equation}
which are the roots of the equation $\lambda=1$. Indeed we had
imposed this additional condition which is unnecessary as it has been
shown in \cite{spagnoli1,spagnoli2} where the more general solution
(\ref{esseparasol},\ref{soluzi}) has been found.
\par
It remains to be seen whether the equivalence between the minimal FDA
$\mathbb{M}$ and the Lie algebra $\widehat{\mathbb{G}}$ can be
promoted to a dynamical equivalence between their gaugings. In other
words can we consistently parametrize the curvatures of
$\widehat{\mathbb{G}}$ in such a way that identifying the three form
$\mathbf{A^{[3]}}$ with the polynomial $\mathbf{\Phi^{[3]}}$, the
rheonomic parametrizations (\ref{rheoFDA}) are automatically
reproduced? This is a rather formidable algebraic problem and to the
present time we have not been able to answer it in the positive or
negative way\footnote{This problem is currently considered in a
collaboration by L. Castellani, P. Fr\'e, F. Gargiulo and K. Rulik
but results are difficult to be obtained mostly because of computer time
limits in the massive algebraic simplifications  which turn out to be needed. It must also
be noted that the algebras defined by equations (\ref{curvacurva}-\ref{curvasigma}) and by some authors
named D'Auria-Fr\'e algebras have been discussed as a possible basis for a Chern-Simons formulation of
fundamental M--theory \cite{nastase,horava}. They have also been retrieved as part of a wider set
of gauge algebras by Castellani \cite{leonardus}, using his method of extended Lie
derivatives.}.
My goal in the present paper is different. I  rather want to
concentrate on the bosonic FDA.s which emerge in twisted tori
compactifications and clarify some issues that were left open in the
current literature on this topic. As I already stressed, my
presentation of the M-theory super FDA had a double purpose. On one
hand I wanted to introduce the structure whose bosonic sector I will
utilize in the analysis of compactifications, on the other hand it
was my purpose to illustrate the working of Sullivan theorems and Chevalley
cohomology in well known examples. The same concepts will be at work
in the proposed object of study.
\section{Compactification of M-theory on Twisted Tori and FDA}
\label{Mtwisted}
If I delete the fermionic generator $\psi$, the FDA of M--theory,
introduced in eq.s (\ref{FDAcompleta}) reduces to the following
system:
\begin{eqnarray}
R^{\hat{a} \hat{b}} & = & d\omega^{\hat{a} \hat{b}} \, -
 \, \omega^{\hat{a} \hat{c}} \, \wedge \, \omega^{\hat{c} \hat{b}}
 \nonumber\\
T^{\hat{a}} & = & \mathcal{D} V^{\hat{a}}\nonumber\\
{\mathbf{F}}^{[4]} & = & d\mathbf{A}^{[3]} \nonumber\\
{\mathbf{F}}^{[7]} & = & d\mathbf{A}^{[6]} -15 \, {\mathbf{F}}^{[4]} \, \wedge \,  \mathbf{A}^{[3]}
\label{boseFDAcomp}
\end{eqnarray}
where I have put a hat on the $D=11$ Lorentz indices
$\hat{a},\hat{b},\dots $. This is just a preparatory step for the
next one, namely compactification on a so called twisted torus. This
is just a fancy name utilized in current literature for a $7$--dimensional
group manifold $\mathcal{G}$ (possibly modded by the action
of some discrete subgroup $\Delta \subset \mathcal{G}$ which makes it compact) whose Lie algebra I assume
to have structure constants $\tau^{I}_{\phantom{I}JK}$ and whose
left--invariant $1$--forms
\begin{equation}
  e^I= e^I_{\mathcal{M}}(y) \, dy^{\mathcal{M}}
\label{linvforme}
\end{equation}
have to be identified with the abstract
forms $e^I$ utilized in Chevalley cohomology (see eq.(\ref{MC1})).
The idea is that of performing the dimensional reduction around the
vacuum:
\begin{equation}
  \mathcal{M}_{11} = \mathcal{M}_4 \, \times \, \mathcal{G}/\Delta
\label{M11split}
\end{equation}
in presence of a flux for the $\mathbf{F^{[4]}}$ field strength and
to analyze the structure of the $D=4$ FDA which emerges from the $D=11$
FDA by this token. To this effect and following the conventions of the recent papers
\cite{RiccaMarioSergioFDA1,RiccaGuidoSergioFDA,curvatureFDA1}, I
adopt the following notations and ans\"atze. The $D=11$ vielbein is
split as follows:
\begin{equation}
  V^{\hat a} =\left \{\begin{array}{rclcrcl}
    V^a & = & E^a & ; & a & = & 0,1,2,3 \\
    V^I & = & e^I+W^I & ; & I & = & 4,5,6,7,8,9,10 \
  \end{array} \right.
\label{conv1}
\end{equation}
Then I consider the spin connection of the $D=11$ theory and I split it as follows:
\begin{equation}
  {\hat \omega}^{{\hat a}{\hat b} } = \left \{ \begin{array}{ccc}
    {\hat \omega}^{ab} & = & 0 \\
    {\hat \omega}^{aI} & = & 0 \\
    {\hat \omega}^{IJ} & = & \omega_{(W)}^{IJ} \, + \, \Delta \omega^{IJ} \
  \end{array}\right.
\label{c1}
\end{equation}
The connection $\omega_{(W)}^{IJ}$ is defined in such a way that
with respect to it there is torsion and the torsion tensor is
precisely related to the structure constants, namely, by definition
we set:
\begin{equation}
  \mathcal{D}_{(W)} V^I = G^I + \ft 12 \,
  \tau^{I}_{\phantom{I}JK}\, V^J \, \wedge \, V^K
\label{t1}
\end{equation}
where
\begin{equation}
  G^I \equiv dW^I +
   \ft 12 \,
  \tau^{I}_{\phantom{I}JK}\, W^J \, \wedge \, W^K
\label{t2}
\end{equation}
is the curvature 2-form of the space-time gauge field $W^I$ gauging
the algebra $\mathbb{G}$.
\par
The connection $\omega_{(W)}^{IJ}$ is completely fixed by the
position (\ref{t1}). Indeed on one hand we must have :
\begin{eqnarray}
0 & = &  \mathcal{D}_{(W)} V^I  - \Delta \omega^{IK} \, \wedge \, V^K \,
  \eta_{JK} \nonumber\\
  & = & \mathcal{D}_{(W)} V^I + \Delta \omega^{IJ} \, \wedge \, V^J \,
\label{t3}
\end{eqnarray}
while on the other hand eq. (\ref{t1}), must be true. The solution is
uniquely given by setting:
\begin{equation}
  \omega^{I}_{(W)\phantom{I}J} = - \tau^{I}_{\phantom{I}JM}\,
  W^M
\label{t8}
\end{equation}
\subsection{Expansion of the $3$-form potential, types of differential forms and their
notations}
\label{3expa}
Next one parametrizes the three form $\mathbf{A}^{[3]}$ of M-theory FDA
according to the chosen compactification, namely decomposes it along
the basis of vielbein (\ref{conv1}), as follows:
\begin{eqnarray}
\mathbf{A}^{[3]} & = & C^{[0]}_{IJK} \, V^I \wedge V^J \wedge V^K + A^{[1]}_{IJ} \wedge V^I \wedge V^J
  \, + \, B^{[2]}_I \, \wedge \, V^I + {A}^{[3]}\nonumber\\
\label{3f1}
\end{eqnarray}
\par
The convention followed in equation (\ref{3f1}) and which I adopt throughout is
that $p$-forms in  $D=11$ space are denoted with boldfaced characters,
$\mathbf{A}^{[p]},\mathbf{B}^{[q]},\mathbf{C}^{[r]},\dots$ and
$\null^{[p]}$ specifies  their degree in such a space. On the other hand
differential forms in $D=4$ are denoted by capital normal letters
${A}^{[p]},{B}^{[q]},{C}^{[r]},\dots$,  $\null^{[p]}$ mentioning
their degree in $D=4$.
\par
Since the algebraic structure underlying all these considerations is
a \textit{double elliptic complex} I will introduce a third type of
differential forms denoted by capital calligraphic letters
${\mathcal{A}}^{[q,p]},{\mathcal{B}}^{[q,p]},{\mathcal{C}}^{[q,p]},\dots$
and characterized not by one, rather by two degrees, one with respect to
$D=4$ space-time and one with respect to the internal group manifold
$\mathcal{G}$. Explicitly we have:
\begin{equation}
  {\mathcal{A}}^{[q,p]} = {A}^{[q]}_{I_1 \dots I_p} \, \wedge e^{I_1}
  \, \wedge \, \dots \,
  \, \wedge \, e^{I_p}
\label{orpello}
\end{equation}
According to these notations the exterior derivative operator $\mathbf{d}$ in
$D=11$ can be rewritten as the sum of two \textit{anticommuting boundary
operators} as it happens in all double elliptic complexes:
\begin{eqnarray}
   \mathbf{d}  & = & d \, + \, \partial \quad \quad ; \quad \mathbf{d}^2 = 0 \nonumber\\
  0 &= & \partial \, d + d \, \partial \quad ;
   \quad d^2 = 0 \quad ; \quad \partial^2 = 0
\label{doubcomplex}
\end{eqnarray}
To complete my set of conventions I also introduce a standard
notation for $p$-forms on the group manifold $\mathcal{G}$. They will
be denoted by capital Greek letters
$\Omega^{[p]},\Sigma^{[q]},\Theta^{[r]},\dots$, so that:
\begin{equation}
  \Omega^{[p]} = \Omega_{I_1 \, \dots \, I_p} \, e^{I_1} \, \wedge \,
  \dots \, e^{I_p}
\label{omeganote}
\end{equation}
When the coefficients $\Omega_{I_1 \, \dots \,
I_p}$ are constant tensors, then $\Omega^{[p]}$ becomes a cochain in
the Chevalley complex for Lie algebra cohomology as summarized in section \ref{chevalley}.
\par
I can now reinterpret the meaning of the  connection $\omega_{(W)}$ introduced in eq.(\ref{t8})
from the point
of view of Chevalley cohomology. Consider a space-time \textit{$q$--form
valued $p$-cochain} of the Chevalley elliptic complex. This means an
object of the  type introduced in eq.(\ref{orpello}).  Following the
so far introduced notations I can calculate the $\mathcal{D}_{(W)}$
derivative of the space-time form ${A}^{[q]}_{I_1 \dots I_p}$ which
plays the role of components for a Chevalley $p$--cochain. Utilizing the definition
of eq.(\ref{t8}) I find
\begin{eqnarray}
 \mathcal{D}_{(W)}{A}^{[q]}_{I_1 \dots I_p}  &=& d{A}^{[q]}_{I_1 \dots
 I_p} \, - \, (-)^{p-1} \, p \, W^M \,
  \tau^{R}_{\phantom{R}M[I_1} \, A^{[q]}_{I_2I_3 \dots I_p]R}
\label{DWdiA}
\end{eqnarray}
Next I consider the new \textit{$(q+1)$--form  valued $p$--cochain} whose components are given by
$\mathcal{D}_{(W)}{A}^{[q]}_{I_1 \dots
I_p}$, namely:
\begin{equation}
  \mathcal{E}^{[q+1,p]} \equiv \mathcal{D}_{(W)}{A}^{[q]}_{I_1 \dots
  I_p}\, e^{I_1} \, \wedge \,  \dots \, \wedge \, e^{I_p}
\label{Dwasform}
\end{equation}
Recalling eq.(\ref{ellXincompo}) I obtain the following
identification:
\begin{equation}
  \mathcal{E}^{[q+1,p]} = d \, {\mathcal{A}}^{[q,p]} -  \, \ell_W \, {\mathcal{A}}^{[q,p]}
\label{DwasLieDer}
\end{equation}
Eq.(\ref{DwasLieDer}) provides a  tool of great help in order
to translate all formulae into an intrinsic notation which exposes
their meaning in Chevalley cohomology. Indeed I can identify the
covariant derivative $\mathcal{D}_{(W)}$ introduced in ref.s \cite{RiccaMarioSergioFDA1,curvatureFDA1} with the following
index-free operator:
\begin{equation}
  \mathcal{D}_{(W)} \equiv d - \, \ell_W
\label{Dwformal}
\end{equation}
which maps $ \mathcal{A}^{[q,p]}$-forms into $
\mathcal{A}^{[q+1,p]}$-forms. Its fundamental property is encoded
into the following identity:
\begin{equation}
  \mathcal{D}_{(W)}^2 = -\ell_G
\label{squareDW}
\end{equation}

\subsection{Expansion of the $4$-form field strength with flux}
\label{4expaflux}
According to the bosonic reduction of M-theory FDA spelled out in eq.s (\ref{boseFDAcomp})
we can calculate the expansion of the $4$--form field strength:
\begin{equation}
  \mathbf{F}^{[4]} = \mathbf{d}\mathbf{A}^{[3]}
\label{F4}
\end{equation}
Following the approach of \cite{RiccaMarioSergioFDA1,curvatureFDA1} I assume that there is
an internal \textit{flux} namely I assume that in the vacuum background
configuration $\mathbf{F}^{[4]}$ is non zero, rather it is equal to
an internal constant $4$--form:
\begin{eqnarray}
  \mathbf{F}^{[4]} &=& \Pi^{[0,4]} \nonumber\\
\Pi^{[0,4]} & = & g_{IJKL} \, e^I \wedge e^J \wedge e^K \wedge
e^K\nonumber\\
g_{IJKL} & = & \mbox{constant tensor}
\label{fluxcondo}
\end{eqnarray}
The Bianchi identity $\mathbf{d}\mathbf{F}^{[4]}=0$ implies that the
form $\Pi^{[0,4]}$ is closed in the Chevalley complex, namely that
it is a Chevalley $4$--cocycle of the Lie algebra $\mathbb{G}$:
\begin{equation}
  \partial \, \Pi^{[0,4]} = 0
\label{3f4}
\end{equation}
In component form eq.(\ref{3f4}) is equivalent to:
\begin{equation}
  \tau^{L}_{\phantom{L}[PQ} \, g_{IJK]L} \, = \, 0
\label{3f4bis}
\end{equation}
Calculating explicitly $\mathbf{dA}^{[3]}$ from the expansion ansatz
(\ref{3f1}) I obtain the following result:
\begin{eqnarray}
{\hat  {\mathbf{F}}}^{[4]} & \equiv & \mathbf{F}^{[4]} -\Pi^{[0,4]} \nonumber\\
\null & = & F^{[4]} \, + \, F^{[3]}_I \,\wedge \,  V^I + \, F^{[2]}_{IJ} \,\wedge \,  V^I \,\wedge
\,V^J \nonumber\\
 && + \, F^{[1]}_{IJK} \,\wedge \,  V^I \,\wedge \,V^J \, \wedge \,
 V^K \, + \,  F^{[0]}_{IJKL} \,\wedge \,  V^I \,\wedge \,V^J \, \wedge \,
 V^K \, \wedge \, V^L
\label{fb1}
\end{eqnarray}
where the curvatures
have the following form identical to that calculated
in \cite{curvatureFDA1}\footnote{The notations used here are identical to those of \cite{curvatureFDA1} with just
an exception. The Kaluza Klein gauge field  coming from the vielbein was named
$A^L$ in \cite{curvatureFDA1}, using the same letter $A$ as used for the gauge fields
coming from the $3$--form. Here we changed notation to  $W^L$, while the gauge fields coming
from the $3$--form remained noted by $A_{\dots}$. This change in notation is not just capricious. The
Chevalley cohomological nature of $W$ is just quite different from that of its relatives $A$. The Kaluza Klein vector
is associated not with forms on the group manifold rather with tangent vectors. It is a section of the tangent
rather than of the cotangent bundle of $\mathcal{G}$ and its role in the Chevalley complex is quite different since
we are supposed to perform Lie derivatives of Chevalley forms along $W$.
For this reason the change in notation was unavoidable. According to this, the covariant derivative
$D^{(\tau)}$ of \cite{curvatureFDA1} has been renamed $D^{(W)}$ mentioning the tangent
vector $W$ out of which it is constructed. The need of
such renaming  is clear when one considers the already discussed transcription (\ref{Dwformal}) of
$D^{(W)}$.} :
\begin{eqnarray}
F^{[0]}_{IJKL} & = & -g_{IJKL} \, + \, \ft 32 \, \tau^{M}_{\phantom{M}[IJ} \, C^{[0]}_{KL]M} \nonumber\\
F^{[1]}_{IJK} & = & \mathcal{D}_{(W)}C_{IJK} \, + \, \tau^{L}_{\phantom{L}[IJ} \, A^{[1]}_{K]L} - 4 \,
g_{IJKL} \, W^L \nonumber\\
 F^{[2]}_{IJ} & = & \mathcal{D}_{(W)}A^{[1]}_{IJ} \, + \,
 \ft 12 \, \tau^{L}_{\phantom{L}IJ} \, B^{[2]}_L \, - \, 6 \, g_{IJLM} \, W^L \, \wedge \, W^M \,
  + \, 3 \,  C^{[0]}_{IJL} \, G^L \nonumber\\
 F^{[3]}_I & = & \mathcal{D}_{(W)}B^{[2]}_{I} \, - \, 2 \, G^J \, \wedge \, A^{[1]}_{JI}
 - 4 \, g_{IJKL} \, W^J \, \wedge \, W^K \, \wedge \, W^L\nonumber\\
F^{[4]} & = & dA^{[3]} \, - \, g_{IJKL} \, W^I \, \wedge \, W^J \, \wedge \, W^K \, \wedge \, W^L \, + \,
 B^{[2]}_I \, \wedge \, G^I
\label{Mariocurve}
\end{eqnarray}
It is now my goal to rewrite formulae (\ref{Mariocurve}) in a more
intrinsic notation that better exposes their  cohomological meaning
within the framework of Chevalley cohomology.
I introduce the definitions:
\begin{equation}
  \begin{array}{|rcl|rcl|}
  \hline
  \mbox{Potentials}&\null&\null&\mbox{Curvatures}&\null&\null\\
  \hline
    \mbox{Flux}^{[0,4]} & = & \Pi^{[0,4]} & \mathcal{F}^{[0,4]} & = &
    F^{[0]}_{IJKL} \, e^I \wedge e^J \wedge e^K \wedge e^L \\
   \mathcal{A}^{[0,3]} & = & C_{IJK} \,  e^I \wedge e^J \wedge e^K  &
   \mathcal{F}^{[1,3]} & = & F^{[1]}_{IJK} \, e^I \wedge e^J \wedge e^K  \\
    \mathcal{A}^{[1,2]} & = & A^{[1]}_{IJ} \,e^I \wedge e^J &
   \mathcal{F}^{[2,2]} & = & F^{[2]}_{IJ} \, e^I \wedge e^J  \\
    \mathcal{A}^{[2,1]} & = & B^{[2]}_I \, e^I & \mathcal{F}^{[3,1]} & = & F^{[3]}_{I} \, e^I  \\
    \mathcal{A}^{[3,0]} & = & A^{[3]} &\mathcal{F}^{[4,0]} & = & F^{[4]}  \\
   %  &  &  &  &  &  \
   \hline
  \end{array}
\label{oppaforme}
\end{equation}
and by means of them I can rewrite eq.s (\ref{Mariocurve}) as
follows:
\begin{eqnarray}
\mathcal{F}^{[0,4]} & = & -\Pi^{[0,4]}\, + \, \partial \, \mathcal{A}^{[0,3]} \nonumber\\
 \mathcal{F}^{[1,3]} & = & (d-\ell_W) \, \mathcal{A}^{[0,3]} \, - \, \partial \,\mathcal{A}^{[1,2]} \, + \,
i_W  \, \Pi^{[0,4]} \nonumber\\
\null & \equiv & \mathcal{D}_{(W)} \, \mathcal{A}^{[0,3]} \, - \, \partial \,\mathcal{A}^{[1,2]} \, + \,
i_W  \, \Pi^{[0,4]} \nonumber\\
 \mathcal{F}^{[2,2]}  & = & (d-\ell_W) \, \mathcal{A}^{[1,2]} + \partial \,  \mathcal{A}^{[2,1]}
\, + \, \ft 12 \, i_W \circ  i_W \, \Pi^{[0,4]} \, + \, i_G \, \mathcal{A}^{[0,3]} \nonumber\\
\null & \equiv & \mathcal{D}_{(W)} \, \mathcal{A}^{[1,2]} + \partial \,  \mathcal{A}^{[2,1]}
\, + \, \ft 12 \, i_W \circ  i_W \, \Pi^{[0,4]} \, + \, i_G \, \mathcal{A}^{[0,3]} \nonumber\\
 \mathcal{F}^{[3,1]}& = & (d-\ell_W) \, \mathcal{A}^{[2,1]} - \, i_G \, \mathcal{A}^{[1,2]}
- \ft 16 \,   i_W \circ  i_W \circ  i_W  \,\Pi^{[0,4]} \nonumber\\
\null & \equiv & \mathcal{D}_{(W)} \, \mathcal{A}^{[2,1]} - \, i_G \, \mathcal{A}^{[1,2]}
- \ft 16 \,   i_W \circ  i_W \circ  i_W  \,\Pi^{[0,4]} \nonumber\\
 \mathcal{F}^{[4,0]} & = & dA^{[3]} - \ft{1}{24} \, i_W \circ  i_W \circ  i_W \circ
 i_W \,\Pi^{[0,4]} \, + \,i_G \, \mathcal{A}^{[2,1]}
\label{rewritto}
\end{eqnarray}
\subsection{Cohomological interpretation of the zero curvature equations and the Minimal FDA.s}
\label{zerocurve}
According to the general structural theory  of FDA.s briefly recalled in the introduction
and in section \ref{sullivano}, the minimal part of any FDA is defined by setting all of its
curvatures to zero, the switching on of curvatures corresponding to
the introduction of contractible generators \cite{comments}.
Furthermore any minimal FDA, in agreement with Sullivan's theorems
should have an interpretation as an extension of a Lie algebra by
means of its cohomology classes. I am therefore primarily interested
in singling out the structure of the minimal algebra emerging from M-theory compactifications on twisted
tori. To this effect I just have to consider the generalized Maurer
Cartan equations which emerge by setting to zero the
$\mathcal{F}^{[p,q]}$ curvatures defined in eq.s(\ref{rewritto}) and
analyze their  implications and cohomological meaning.
\par
Let us begin with the first equation $\mathcal{F}^{[0,4]}=0$. This
implies that the flux  $\Pi^{[0,4]}$ is actually an exact form:
\begin{equation}
  \Pi^{[0,4]}= \partial \, \mathcal{A}^{[0,3]}
\label{1zerocondo}
\end{equation}
Furthermore since $\Pi^{[0,4]}$ is constant it also implies that the
antisymmetric tensor $C_{IJK}$ is constant as well:
\begin{equation}
  d \mathcal{A}^{[0,3]}=0
\label{0Ccondo}
\end{equation}
Consider next the second equation $\mathcal{F}^{[1,3]}=0$.
Substituting into it the previous result (\ref{1zerocondo}) we
obtain:
\begin{eqnarray}
0 & = & \left( d -\ell_W\right) \, \mathcal{A}^{[0,3]}\, - \, \partial \mathcal{A}^{[1,2]} \, + \, i_W \,
\partial \mathcal{A}^{[0,3]} \nonumber\\
\null & = &\partial \left( \mathcal{A}^{[1,2]} \, + \, i_W \,
\mathcal{A}^{[0,3]} \right)
\label{A12condo}
\end{eqnarray}
The general solution of eq.(\ref{A12condo}) is given by:
\begin{equation}
  \mathcal{A}^{[1,2]} = \Sigma^{[1,2]} \, - \, i_W\, \mathcal{A}^{[0,3]}
\label{A12soluz}
\end{equation}
where $\Sigma^{[1,2]}$ is a $1$--form valued cocycle of Chevalley
cohomology. Therefore, using the standard basis introduced in table
\ref{unciricamo}, we can write the expansion:
\begin{eqnarray}
 \Sigma^{[1,2]} &=& \Sigma^{[1,2]}_\perp \, \oplus \,
  \Sigma^{[1,2]}_\| \nonumber\\
  \Sigma^{[1,2]}_\perp & = & \sum_{a=1}^{h_2} \,Z_{[1]}^a \, \Gamma^{[2]}_a \nonumber\\
  \Sigma^{[1,2]}_\| & = & \partial \Upsilon^{[1,1]} \nonumber\\
  \Upsilon^{[1,1]} & = & \sum_{\bar{\alpha}=1}^{7-h_1} \,\Upsilon_{[1]}^{\bar{\alpha}} \,
  \Xi^{[1]}_{\bar{\alpha}}
\label{sig12}
\end{eqnarray}
where $Z_{[1]}^a$ are $h_2$ new gauge $1$--forms in $D=4$ and
similarly $\Upsilon_{[1]}^{\bar{\alpha}}$ are other $7-h_1$ such
gauge fields.
\par
Let us now analyze the third equation $\mathcal{F}^{[2,2]}=0$.
Inserting into it the previous results we obtain:
\begin{eqnarray}
0 & = & \left(d - \ell_W \right) \, \Sigma^{[1,2]} \, - \,  \left(d - \ell_W \right) \, i_W \mathcal{A}^{[0,3]}
%\nonumber\\
%\null & \null&
+ \ft 12 \, i_W\circ i_W\, \partial\mathcal{A}^{[0,3]} \, + \, i_G \mathcal{A}^{[0,3]} \, + \,
\partial \mathcal{A}^{[2,1]}
\label{B12equa}
\end{eqnarray}
By formal manipulations, using the definition of the Lie derivative (\ref{ellXdefi}) and
the definition of the contraction operation (\ref{idiX}),
eq.(\ref{B12equa}) can be rewritten as follows:
\begin{equation}
  0 = d \Sigma^{[1,2]} \, - \, \partial \left( i_W \Sigma^{[1,2]} - \ft 12 \, i_W\circ i_W \mathcal{A}^{[0,3]} \, - \,
  \mathcal{A}^{[2,1]}\right) \, + \, Q^{[2,2]}
\label{dsig12}
\end{equation}
where the remaining form $Q^{[2,2]}$ defined below is actually
identically zero:
\begin{equation}
  Q^{[2,2]} \, \equiv \, i_G \, \mathcal{A}^{[0,3]} \, - \, i_{dW} \,
  \mathcal{A}^{[0,3]} \, + \, \ft 12 \, \left( \ell_W \circ i_W \, + \, i_W  \circ \ell_W
  \right) \mathcal{A}^{[0,3]}\, = \, 0
\label{q22equa}
\end{equation}
Eq.(\ref{q22equa}) follows immediately from the definition of the
curvature $G$ in eq. (\ref{t2}) and from the identity:
\begin{equation}
  -i_{\left[ W \, ,\, W\right ]} \mathcal{C}^{[p]} = \left( i_W \circ \ell_W +\ell_W
  \circ i_W \right) \, \mathcal{C}^{[p]}
\label{ideWW}
\end{equation}
holding true for any $p$--cochain and following from eq.
(\ref{LxiY}). In order to interpret equation (\ref{dsig12}), we just
have to recall the orthogonal decomposition (\ref{sig12})
to separate the form  $\Sigma^{[1,2]}$. into its trivial and non
trivial part. In this way, equation (\ref{dsig12}) can be decomposed in two
equations:
\begin{eqnarray}
0 & = & d\Sigma^{[1,2]}_\perp  \label{contra1}\\
0 & = &\partial \left( d\Upsilon^{[1,1]} + i_W \Sigma^{[1,2]} - \ft 12 \, i_W\circ i_W \mathcal{A}^{[0,3]} \, - \,
  \mathcal{A}^{[2,1]}\right) \label{noncontra2}
\end{eqnarray}
Equation (\ref{contra1}) simply states that the gauge--fields
$Z_{[1]}^{x}$ associated with the $h_2$ harmonic
$2$--form of $\mathbb{G}$ are abelian and just contribute a contractible
sector to the $D=4$ FDA:
\begin{equation}
 d {Z}_{[1]}^{x} = 0
\label{h2contractible}
\end{equation}
Eq.(\ref{noncontra2}) needs instead a little more elaboration. I
have to recall that, for $1$--cycles $C^{[1,1]}$, the Lie derivative
(\ref{ellXdefi}) is just given by the first term only:
\begin{equation}
  \ell_W \, C^{[1,1]} = i_W\circ \partial\, C^{[1,1]}
\label{ellWonY}
\end{equation}
The reason is that on $0$-chains the Chevalley boundary operator
$\partial$ is identically zero and hence $ \partial\, i_WC^{[1,1]}
\equiv 0$. Taking this observation into account equation
(\ref{noncontra2}) is solved by setting:
\begin{eqnarray}
  \mathcal{A}^{[2,1]} &=& \Sigma^{[2,1]}_\perp \, + d\Upsilon^{[1,1]} \, + \, i_W \, \Sigma^{[1,2]} \,
  - \, \ft 12 \, i_W\circ i_W
  \,\mathcal{A}^{[0,3]} \nonumber\\
  & = & \Sigma^{[2,1]}_\perp \, + \, \mathcal{D}^{W}
  \Upsilon^{[1,1]} \, + \, i_W \, \Sigma^{[1,2]}_\perp \, - \, \ft 12 \, i_W\circ i_W
  \,\mathcal{A}^{[0,3]}
\label{B12}
\end{eqnarray}
where $\Sigma^{[2,1]}_\perp$ is a $2$--form valued Chevalley $1$-cocycle:
\begin{equation}
  \partial \Sigma^{[2,1]}_\perp = 0
\label{Gammacycle}
\end{equation}
Recalling our basis conventions in table \ref{unciricamo} we can
write:
\begin{equation}
  \Sigma^{[2,1]}_\perp = \sum_{\alpha=1}^{h_1} \, B^\alpha_{[2]} \,
  \Gamma^{[1]}_\alpha
\label{decompoG21}
\end{equation}
\par
Next I analyze the implications of the equation
$\mathcal{F}^{[3,1]}=0$. Inserting eq.(\ref{A12soluz}) and the first
of eq.s (\ref{B12}) into the definition of $\mathcal{F}^{[3,1]}$ (see
eq.s (\ref{rewritto})) I obtain four type of contributions:
\begin{equation}
  \mathcal{F}^{[3,1]}= \mathcal{T}(\Sigma^{[2,1]}_\perp)\, + \, \mathcal{T}(\Sigma^{[1,2]}_\perp)\,
  + \,  \mathcal{T}(\Upsilon^{[1,1]})
  \, + \,  \mathcal{T}(\mathcal{A}^{[0,3]})
\label{trecontributi}
\end{equation}
those in $\Sigma^{[2,1]}_\perp$, those in $\Sigma^{[1,2]}_\perp$, those in $\Upsilon^{[1,1]}$
and those in $\mathcal{A}^{[0,3]}$, respectively.
Let me elaborate them separately. I begin with $\mathcal{T}(\Sigma^{[2,1]}_\perp)$ and I find:
\begin{equation}
  \mathcal{T}(\Sigma^{[2,1]}_\perp) = \left(d - \ell_W \right) \Sigma^{[2,1]}_\perp =
\left(d - \partial \, i_W \right) \Sigma^{[2,1]}_\perp = d \Sigma^{[2,1]}_\perp
\label{oppiGamma}
\end{equation}
The first step in (\ref{oppiGamma}) follows since $\partial
\Sigma^{[2,1]}_\perp=0$, the third because $i_W \Sigma^{[2,1]}_\perp$ is a
$0$--cochain in Chevalley cohomology. Next I consider the $\mathcal{T}\left(
\Sigma_\perp^{[1,2]}\right)$ contribution. Here I utilize the identity:
\begin{equation}
  -\ell_W \circ i_W \, \Sigma^{[1,2]}_\perp = \ft 12 i_{\left[ W \, , \, W \right]
  } \, \Sigma^{[1,2]}_\perp
\label{ellWiWonSigma}
\end{equation}
which follows since $\Sigma^{[1,2]}_\perp$ is a cocycle
$\partial\Sigma^{[1,2]}_\perp=0$. By means of this identity we can
evaluate:
\begin{eqnarray}
  \mathcal{T}(\Sigma_\perp^{[1,2]}) & = & \left( d - \ell_W\right)  \, i_W
  \Sigma^{[1,2]}_\perp - i_G \Sigma^{[1,2]}_\perp \nonumber\\
  & = & -i_W d \Sigma^{[1,2]}_\perp
\label{Tsigmaelabo}
\end{eqnarray}
Let us now consider the terms in $\Upsilon^{[1,1]}$. Here we have:
\begin{equation}
  \mathcal{T}(\Upsilon^{[1,1]})= \mathcal{D}_{(W)}^2\,\Upsilon^{[1,1]} - i_G
  \Upsilon^{[1,1]} = 0
\label{Tupselabo}
\end{equation}
Finally let us consider the terms $\mathcal{T}(\mathcal{A}^{[0,3]})$. Here we
have:
\begin{eqnarray}
\mathcal{T}(\mathcal{A}^{[0,3]}) & = & \mathcal{D}_{(W)}\left(- \ft 12 i_W\circ i_W \mathcal{A}^{[0,3]} \right) \, + \,
i_G \circ i_W \mathcal{A}^{[0,3]} \, - \, \ft 16 i_W \circ i_W \circ i_W \partial \mathcal{A}^{[0,3]} \nonumber\\
\null & = & \ft 12  i_W \circ \ell_W \circ i_W \, \mathcal{A}^{[0,3]}
 - \ft 16 i_W \circ i_W \circ i_W \partial \mathcal{A}^{[0,3]}
 \nonumber\\
 \null & = & 0
\label{omniaCvanish}
\end{eqnarray}
The last line of eq.(\ref{omniaCvanish}) follows from an explicit
evaluation of the two terms in the second line.
\paragraph{The Minimal FDA in $D=4$}
We can now summarize our results. The $0$--curvature equations
\begin{equation}
 \mathcal{F}^{[p,q]} = 0
\label{omniF}
\end{equation}
lead to the following $D=4$ bosonic free differential algebra.
Besides the seven $W^I$ one--forms gauging the original algebra
$\mathbb{G}$, there are $h_3$ zero-forms $\Phi_{[0]}^x$,  $h_2$ additional gauge one--forms
$Z_{[1]}^a$ and $h_1$ gauge two--forms $B_{[2]}^\alpha$ and
$h_0 = 1$ three-forms
\begin{equation}
  \overline{A}^\ell_{[3]} \equiv {A}^{[3]} -C_{IJK} \, W^{I} \wedge W^J
  \wedge W^K
\label{redefinoA3}
\end{equation}
 closing the algebra:
\begin{equation}
\begin{array}{||rcccl||}
\hline
\hline
  dW^I + \ft 12 \tau^{I}_{\phantom{I}JK} \, W^J \, \wedge \, W^K & = & 0 & ; & I=1,\dots , \mbox{dim} \, \mathbb{G} \\
  d\Phi^x_{[0]} & = & 0 & ; & x=1,\dots , h_{3} (\mathbb{G}) \\
  dZ_{[1]}^a & = & 0 & ; & a=1,\dots, h_2(\mathbb{G}) \\
  dB_{[2]}^{\alpha} & = & 0 & ; & {\alpha}=1,\dots, h_1(\mathbb{G}) \\
   d\overline{A}^\ell_{[3]} & = & 0  & ; & \ell = 1,\dots , h_0(\mathbb{G})=1 \\
  \hline
  \hline
\end{array}
\label{FDAquantum}
\end{equation}
where $h_{0,1,2,3}$  are the dimensions of the
cohomology groups of $\mathbb{G}$. In addition there are also $\wp_2$
gauge fields $\,\Upsilon_{[1]}^{\bar{a}} \,$ on which no
condition is imposed, where $\wp_2=7-h_1$ is the dimension of $\mbox{Im}
\partial_1$. These gauge fields are non physical, they are the gauge
degrees of freedom of the $2$-forms. It is evident from their
structure that the minimal part of the FDA is simply given by the
original algebra $\mathbb{G}$ which is a standard Lie algebra,  the
remaining three equations, defining instead contractible generators.
\par
With reference to the D=4 FDA summarized in eq.(\ref{FDAquantum}) we
can now summarize the complete parametrization of the original potentials $
\mathcal{A}^{[p,q]}$ displayed in table (\ref{oppaforme}). The
independent objects are the following five:
\begin{eqnarray}
\Sigma^{[0,3]}_\perp &=&\sum_{x=1}^{h_3} \,  \Phi^x_{[0]}  \,
\Gamma^{[3]}_x \nonumber\\
\Sigma^{[1,2]}_\perp &=& \sum_{a=1}^{h_2} \, Z^a_{[1]} \,
  \Gamma^{[2]}_a \nonumber\\
\Sigma^{[2,1]}_\perp &=& \sum_{\alpha=1}^{h_1} \, B^\alpha_{[2]} \,
  \Gamma^{[1]}_\alpha\nonumber\\
  \Sigma^{[3,0]}_\perp & = & \sum_{\ell = 1}^{h_0=1} \overline{A}_{[3]}^\ell \,
  \Gamma^{[0]}_\ell
  \nonumber\\
  \Upsilon^{[0,3]} & = &  \sum_{\bar{x}=1}^{21-h_3+h_2-h_1} \,  \phi^{\bar{x}}  \,  \Xi^{[3]}_{\bar{x}} \, \nonumber\\
\Upsilon^{[0,2]} & = & \sum_{\bar{a}=1}^{14-h_2+h_1} \,  \phi^{\bar{a}}  \, \Xi^{[2]}_{\bar{a}} \nonumber\\
  \Upsilon^{[1,1]} & = & \sum_{\bar{\alpha}=1}^{7-h_1} \,\Upsilon_{[1]}^{\bar{\alpha}} \,
  \Xi^{[1]}_{\bar{\alpha}}
\label{independenceday}
\end{eqnarray}
where  $\phi^{\bar{a}} $, $\phi^{\bar{x}} $ are $35-h_3$ scalar fields coming from the
M--theory three--form subdivided into the two sectors of $3$--cochains orthogonal to the harmonic cycles.
 Their vev.s are arbitrary numbers at the present level
and they can be seen as \textit{moduli} of the algebra. Similarly, as I
have already stressed, the $7-h_1$   one--forms  $\Upsilon_{[1]}^{\bar{\alpha}}$ are
just gauges.
\par
In terms of the objects defined in equations (\ref{independenceday})
the potentials $\mathcal{A}^{[p,q]}$ are explicitly parametrized as follows and this is the most
general solution of the zero curvature equations:
\begin{eqnarray}
\Pi^{[0,4]} & = & \partial\, \Upsilon^{[0,3]}  \nonumber\\
\mathcal{A}^{[0,3]} & = & \Sigma^{[0,3]}_\perp \, + \,
\Upsilon^{[0,3]} \, + \, \partial \, \Upsilon^{[0,2]}\nonumber\\
\mathcal{A}^{[1,2]} & = & \Sigma^{[1,2]}_\perp \, + \, \partial \,
\Upsilon^{[1,1]} \, - \, i_W\, \Sigma^{[0,3]}_\perp \, - \, i_W\, \Upsilon^{[0,3]}\nonumber\\
\mathcal{A}^{[2,1]} & = & \Sigma^{[2,1]}_\perp \, + \, \mathcal{D}^W
\Upsilon^{[1,1]} \, + \, i_W\, \Sigma^{[1,2]}_\perp \, - \, \ft 12 \,
i_W \circ i_W \Sigma^{[0,3]}_\perp \, - \, \ft 12 \,
i_W \circ i_W \Upsilon^{[0,3]}_\perp\nonumber\\
\mathcal{A}^{[3,0]} & = &\Sigma^{[3,0]}_\perp + \ft 1 6 i_W\circ i_W
\circ i_W \Sigma^{[0,3]}_\perp + \ft 1 6 i_W\circ i_W
\circ i_W \Upsilon^{[0,3]}
\label{misurato}
\end{eqnarray}
\par
\paragraph{Conclusive Remarks on the Minimal FDA}
Equations (\ref{FDAquantum}), (\ref{independenceday}) and (\ref{misurato}) are
written in a rather pedantic way, but this is done in order to stress the
conceptual implications of the zero curvature equations. As we have
already remarked, according to the general theory of FDA.s (see section \ref{sullivano} )
the new generators of the algebra should be in one--to--one
correspondence with the cohomology classes of the original algebra
and indeed they are. Yet the correspondence is that anticipated in
eq.(\ref{Mpredi}) and not that involved in Sullivan's second theorem,
namely that of eq.(\ref{Sullivanpredi}). Indeed
the number $h_0$ which is just trivially equal to one
predicts the number of $D=4$ three--forms, the number $h_1$ predicts
the number of two--forms, the number $h_2$ predicts the number of
additional one--forms. Finally we expect that $h_3$ should predict
the number of essential $0$--forms, or scalars.  At this level we do
not see it clearly, since all the $35$ scalars are any how constants,
both those associated with non trivial $h_3$--classes and those
associated with boundaries. The vev.s of all the scalars are to be
regarded as moduli. In any case the D=4 FDA defined by the zero
curvature equations  is a trivial FDA since all new generators are
contractible, as we have already remarked.
\subsection{Introducing the FDA curvatures and non trivial fluxes}
\label{introcurve}
It is now fairly easy to deform the minimal FDA by introducing curvatures
and, as I shall emphasize, it is only in presence of these latter that the
flux $\Pi^{[0,4]}$ can be cohomologically non trivial.
So let us introduce the curvatures of degree $[0,4]$ by setting:
\begin{eqnarray}
 \mathcal{F}^{[0,4]} & = & \mathcal{R}^{[0,4]}_\perp \, + \,  \mathcal{R}^{[0,4]}
  _\partial \nonumber\\
\mathcal{R}^{[0,4]}
  _\partial & = &\partial Q^{(3)} \left( \Pi^{[0,4]}\right)  - \partial \, \mathcal{A}^{[0,3]}\nonumber\\
   \mathcal{R}^{[0,4]}_\perp  & = & P^{(4)}_\perp \left(
   \Pi^{[0,4]}\right)\, \equiv \,
  \Delta^{[0,4]}  =  \sum_{x=1}^{h_4} \, \mu^x \, \Gamma^{[4]}_x
\label{plussone}
\end{eqnarray}
where $\mu^x$ are $h_4=h_3$ constant parameters.
 The crucial
observation is that the zero curvature equations I have
previously solved are modified by the addition to each curvature $\mathcal{F}^{[p,q]}$ of terms of the following form:
\begin{equation}
\begin{array}{ccccl}
  \mathcal{F}^{[1,3]} & \rightarrow &  \mathcal{F}^{[1,3]} & + & i_W \, \Delta^{[0,4]}\, \\
  \mathcal{F}^{[2,2]} & \rightarrow &  \mathcal{F}^{[2,2]} & + &\ft 12 i_W \circ i_W  \, \Delta^{[0,4]}\, \\
   \mathcal{F}^{[3,1]} & \rightarrow &  \mathcal{F}^{[3,1]} & - & \ft 1 6 i_W \circ i_W \circ i_W\, \Delta^{[0,4]}\, \\
   \mathcal{F}^{[4,0]} & \rightarrow &  \mathcal{F}^{[4,0]} & - &
   \ft {1} {24} i_W \circ i_W \circ i_W\, \, \Delta^{[0,4]}\, \\
\end{array}
\label{giannone}
\end{equation}
Let us see the implications of these additions. Modifying eq.(\ref{A12condo}) I obtain:
\begin{eqnarray}
\mathcal{F}^{[1,3]} & = & \left( d -\ell_W\right) \, \mathcal{A}^{[0,3]}\, - \, \partial \mathcal{A}^{[1,2]} \, + \, i_W \,
\partial \mathcal{A}^{[0,3]} + i_W \, \Delta^{[0,4]} \nonumber\\
\null & = & d\mathcal{A}^{[0,3]}\, + \, \partial \left( \mathcal{A}^{[1,2]} \, + \, i_W \,
\mathcal{A}^{[0,3]} \right) + i_W \, \Delta^{[0,4]}
\label{A12condo.bis}
\end{eqnarray}
The $3$--cochain $i_W \, \Delta^{[0,4]}$ has generically a non trivial
projection along all three subspaces $\mathbf{\Gamma}^{[3]}$, $\partial
\mathbf{\Xi}^{[2]}$ and $\mathbf{\Xi}^{[3]}$ (see
eq.s (\ref{decompo3}) and (\ref{projettori})). Defining:
\begin{eqnarray}
\mathcal{R}^{[1,3]}_\perp & = & d\Sigma^{[0,3]}\, + \, P^{(3)}_\perp \left(i_W \, \Delta^{[0,4]} \right)  \nonumber\\
\mathcal{R}^{[1,3]}_\| & = & d\Upsilon^{[0,3]} \, + \, P^{(3)}_\| \left(i_W \, \Delta^{[0,4]}
\right) \nonumber\\
\mathcal{R}^{[1,3]}_\partial & = &  \partial \left[-d\Upsilon^{[0,2]}
- \, i_W\, \Sigma^{[0,3]}_\perp \, - \, i_W\, \Upsilon^{[0,3]} \, + \,
Q^{(2)}_\| \left(i_W \, \Delta^{[0,4]} \right) \right]
\label{R13}
\end{eqnarray}
the curvature $\mathcal{F}^{[1,3]}$ is decomposed into its three independent  projections.
\begin{equation}
  \mathcal{F}^{[1,3]} = \mathcal{R}^{[1,3]}_\perp \, + \,
  \mathcal{R}^{[1,3]}_\| \, + \, \mathcal{R}^{[1,3]}_\partial
\label{office13}
\end{equation}
In a similar way I can decompose all the other curvatures. For
instance let me focus on the curvature $\mathcal{F}^{[3,1]}$ and consider
the transverse component $\mathcal{R}^{[1,3]}_\perp$. Here I find:
\begin{equation}
  \mathcal{R}^{[3,1]}_\perp = d\Sigma_\perp^{[2,1]} \, + \,
  \underbrace{P^{(1)}_\perp \left(-i_W \,d\Sigma^{[1,2]} \, - \,\ft 12 i_W\circ i_W d\mathcal{A}^{[0,3]}
  \right)}_{\mbox{curvature terms}} \, + \, \underbrace{P^{(1)}_\perp \left(- \ft 1 6 \, i_W\circ i_W \circ
  i_W \, \Delta^{[0,4]} \right)}_{\mbox{const term}}
\label{ominous}
\end{equation}
In the above equation as in the first of eq.s (\ref{R13}) you can see the conditions under which a non trivial
FDA could emerge from flux compactifications on twisted tori. All
terms in (\ref{ominous}) are derivative terms, except the last one
which is defined by the non trivial flux $\Delta^{[0,4]}$.
Recalling eq.(\ref{projettori}), from (\ref{ominous}) we extract:
\begin{equation}
F[B]^\alpha =  dB^{\alpha}_{[2]} \, - \, \ft 1 6 \, \underbrace{\langle i_W\circ i_W \circ
  i_W \, \Delta^{[0,4]} \, , \, \Gamma_\alpha^{[6]} \rangle}_{\Lambda^{[3]}_{\alpha}(W)} \, +
  \, \mbox{ derivative terms of lower degree forms}
\label{urca1}
\end{equation}
The object
\begin{equation}
  \Lambda^{[3]}_\alpha(W) =\langle i_W\circ i_W \circ
  i_W \, \Delta^{[0,4]} \, , \, \Gamma_\alpha^{[6]} \rangle
\label{bellet}
\end{equation}
is  a cubic polynomial in the  $1$--forms $W^I$. According to Sullivan's second theorem it is necessarily a
cohomology class of $\mathbb{G}$, or zero.
\par
The condition for non trivial deformations is  therefore  that
$\Lambda^{[3]}_\alpha(W)$ should be non zero or that
\begin{equation}
  \Lambda^{[2]}_a(W) =\langle i_W \circ
  i_W \, \Delta^{[0,4]} \, , \, \Gamma_a^{[5]} \rangle
\label{bellet2}
\end{equation}
should be non zero. When $\Lambda^{[3]}_\alpha(W) \ne 0$ we have  non
trivial $2$--form generators. On the other hand when
$\Lambda^{[2]}_a(W)\ne 0$ we have non trivial $1$--form generators,
namely the Lie algebra $\mathbb{G}$ is extended to some bigger
algebra.
\par
Let us now immediately verify that $ \Lambda^{[3]}_\alpha(W)=0$ in
the case of SS algebras with all different eigenvalues $0\ne m_1 \ne m_2 \ne m_3$. In this case
the flux $\Delta^{[0,4]}$ is a linear combination of the forms
$\Gamma^{[4]}_x$ presented in eq.(\ref{lubiapitni}). Correspondingly
we have:
\begin{equation}
  i_W\circ i_W \circ
  i_W \, \Delta^{[0,4]} \propto   W^{i_1} \, \wedge \, W^{i_2} \, \wedge \, W^{i_3}
  \, \wedge \, e^m \, \mu^x \, \epsilon_{i_1i_2i_3muv} \, \sigma^{uv}_x \,
\label{ielena}
\end{equation}
The form $\Gamma^{[6]}$ is instead given in eq.(\ref{gamma1gamma6})
and it is immediately evident that the wedge product $i_W\circ i_W \circ
i_W \, \Delta^{[0,4]} \wedge \Gamma^{[6]} $ is zero since
$\Gamma^{[6]}$ contains already all the six $e^i$, so that
necessarily there are in the wedge product two identical $e^j$.
Consider instead $\Lambda^{[2]}_a(W)$ for the
same case of SS algebra. Here things are different. On one hand we have:
\begin{equation}
 i_W \circ
i_W \, \Delta^{[0,4]}\propto W^{i_1} \, \wedge \, W^{i_2} \, \wedge \, e^{m}
  \, \wedge \, e^n \, \mu^x \, \epsilon_{i_1i_2mnuv} \, \sigma^{uv}_x \,
\label{dueWW}
\end{equation}
on the other hand, $\Gamma^{[5]}_b$, as given in
eq.(\ref{tentenna}), contains $e^0$ and four $e^i$. It follows
that $\Lambda^{[2]}_a(W)$ is non-zero rather it is of the form:
\begin{equation}
  \Lambda^{[2]}_a(W) \propto \mu^x \, \epsilon_{i_1i_2 mn uv} \,
  \sigma^{mn}_a \, \sigma^{uv}_x \, W^{i_1} \, W^{i_2}
\label{deformatini}
\end{equation}
It means that we have an enlargement of the $\mathbb{G}$ Lie algebra induced
by the non trivial flux. It is interesting to check how Sullivan
theorem is anyhow verified. According to Sullivan's second theorem
the $2$--form $\Lambda^{[2]}_a(W)$ should be a harmonic $2$--form of
$\mathbb{G}$, namely should be a linear combination of $\Gamma^{[2]}_a$ as
displayed in eq.(\ref{tentenna}) with $e^I$, replaced by $W^I$.
Indeed it is! To see it it suffices to recall that, by definition, the
matrices $\sigma_{1,2,3}$ are $\mathrm{SO(6)}$ Cartan generators, namely they are skew
diagonal:
\begin{eqnarray}
 \mu^x \, \sigma_x & = &  \left(\begin{array}{cc|cc|cc}
   0 & \mu_1 & 0 & 0 & 0 & 0 \\
   -\mu_1 & 0 & 0 & 0 & 0 & 0 \\
   \hline
   0 & 0 & 0 & \mu_2 & 0 & 0 \\
   0 & 0 & -\mu_2 & 0 & 0 & 0 \\
   \hline
   0 & 0 & 0 & 0 & 0 & \mu_3 \\
   0 & 0 & 0 & 0 & -\mu_3 & 0 \
 \end{array} \right)
\label{herpes}
\end{eqnarray}
As a consequence of this  $\Lambda^{[2]}_a(W)$ is of the form:
\begin{eqnarray}
  \Lambda^{[2]}_1(W) & \propto & \mu_2 \, \sigma_3^{uv} \,W^u \wedge W^v
  + \mu_3 \, \sigma_2^{uv} \,W^u \wedge W^v \nonumber\\
   \Lambda^{[2]}_2(W) & \propto & \mu_1 \, \sigma_3^{uv} \,W^u \wedge W^v
  + \mu_3 \, \sigma_1^{uv} \,W^u \wedge W^v \nonumber\\
 \Lambda^{[2]}_3(W) & \propto & \mu_1 \, \sigma_2^{uv} \,W^u \wedge W^v
  + \mu_2 \, \sigma_1^{uv} \,W^u \wedge W^v \nonumber\\
  \label{tuttobene}
\end{eqnarray}
which is precisely what it is required by Sullivan's theorem!
\par
Let me finally discuss how also the Maurer Cartan equations of the
$2$--forms can become non trivial in presence of a non trivial
$4$--form flux, if the SS algebra is degenerate. To this effect let
me consider the case $m_1 \ne 0$, $m_2=m_3=0$. Here the fourth
cohomology group is larger and we also have harmonic $4$--forms of
the type:
\begin{equation}
  \Gamma^{(4)} \propto e^0 \wedge e^{i_1} \wedge e^{i_2} \wedge
  e^{i_3} \, U_{i_1i_2i_3}
\label{perdinci}
\end{equation}
provided the tensor $U_{ijk}$ vanishes when one of its indices takes
the values $1$ or $2$. If we use such a harmonic form in the
construction of $\Delta^{[0,4]}$, namely of the flux, then
$\Lambda^{[3]}_x(W)$ as defined by equation (\ref{bellet}) can
receive non vanishing contributions and we can have a genuine
deformation also of the $2$--form Maurer Cartan equations.
\section{Conclusions and Perspectives}
\label{concludo}
In this paper I have recast the analysis of Free Differential
Algebras emerging from M--theory compactifications on so called
twisted tori, into the language of Chevalley cohomology and within
the framework of the structural theorems on FDA.s due to Sullivan. The
main results that I have obtained are:
\begin{enumerate}
  \item Establishing a general and compact double elliptic complex formalism
  which enables the analysis of all explicit cases within the
  cohomology framework of Lie algebras. In particular I have
  established general formulae for the calculation of the spectrum of
  $p$--forms in terms of Hodge numbers of the Lie algebra
  $\mathbb{G}$ and of the generalized structure constants in terms of
  Poincar\'e pairing on the Chevalley complex.
  \item Explicit calculation of the cohomology for the simplest SS
  algebras and construction of an explicit orthogonal basis for the
  $p$--cochain spaces.
  \item A general cohomological analysis of the zero curvature
  equations which reveals how the minimal FDA is always trivial if
  the $4$--form flux is cohomologically trivial
  \item An explicit algebraic criterion to be satisfied by the
  $4$--form flux in order to generate non trivial FDA.s
\end{enumerate}
The applications and future development lines of the presented
results are several. In particular it is worth mentioning:
\begin{enumerate}
  \item Analysis of a wider class of algebras $\mathbb{G}$ within the
  presented framework and a systematic search on the structure of FDA
  obtained thereof, in relation with the available $4$--form fluxes.
  \item Revisited analysis of the dual gauge algebras in relation
  with the presented cohomological set up.
  \item Extension of the present analysis from M--theory to Type IIB
  compactifications.
  \item Use of twisted tori backgrounds and of the formalism presented here
   as a case study in relation with the open problem of
  gauging  M-theory algebras (\ref{curvacurva}-\ref{curvasigma})
  \cite{leonardus}
\end{enumerate}
Work on these issues is on agenda.


\newpage
\begin{thebibliography}{99}
\bibitem{fluxes0} For an exhaustive review on flux compactifications
see A.R. Frey \emph{Warped Strings: Self dual flux and contemporary
compactifications} hep-th/0308156.
\bibitem{fluxes1} S. Gukov, \emph{Solitons, superpotentials and
calibrations} Nucl. Phys. {\bf B574} (2000) 169 [hep-th/9911011]
S. Gukov, C. Vafa and E. Witten Nucl. Phys. {\bf B584} (200) 69
[hep-th/9906070]
\bibitem{fluxes2} G.L. Cardoso, G. Dall'Agata, D. Luest, P.
Manousselis and G. Zoupanos \emph{Non Kaehler string backgrounds and
their five torsion classes} Nucl. Phys. {\bf B652} (2003) 5
[hep-th/0211118].
G.L. Cardoso, G. Curio, G. Dall'Agata and D. Luest, \emph{BPS action
and superpotential for heterotic string compactifications with
fluxes} JHEP \textbf{10} (2003) 004 [hep-th/0306088]
G.L. Cardoso, G. Curio, G. Dall'Agata and D. Luest, \emph{Heterotic
string theory on non--kaeler manifolds with H-flux and gauguno
condensate} Fortsch. Phys. \textbf{52} (2004) 483 [hep-th/03100021].
\bibitem{Ssmecha} J. Sherk and J.H. Schwarz Phys. Lett. {\bf B82} (1979)
60, J. Sherk and J.H. Schwarz \emph{How to get masses form extra
dimensions} Nucl. Phys. {\bf B153} (1979) 61
\bibitem{Hullo} C.M. Hull, R.A. Reid-Edwards \emph{Flux
Compactifications of String Theory on Twisted Tori} hep-th/0503114.
\bibitem{oddstory} N. Kaloper and R.C. Myers \emph{The O(dd) story of
massive supergravity} JHEP \textbf{05} (1999) 010 [hep-th/9901045]
\bibitem{triveditripati} S. Kachru, M.B. Schulz, P.K. Tripathy and
S.P. Trivedi \emph{New supersymmetric string compactifications} JHEP
\textbf{03} (2003) 061 [hep-th/0211182].
\bibitem{fluxesschulz} M.B. Schulz \emph{Superstring orientifolds
with torsion: O5 orientifolds of torus fibrations and their massless spectra}
Fortsch. Phys. \textbf{52} (2004) 963 [hep-th/0406001].
\bibitem{derendin} J.P. Derendinger, C. Kounnas, P.M. Petropoulos, F.
Zwirner, \emph{Superpotentials in IIA compactifications with fluxes}
hep-th/0411276
%
\bibitem{gaugedsugrapot1} F. Cordaro, P. Fr\`e, L. Gualtieri, P. Termonia, M.Trigiante,
\emph{N=8 gaugings revisited: an exhaustive classification},
Nucl.Phys. B532 (1998) 245-279, hep-th/9804056;
%
\bibitem{gaugesugrapot2} L. Andrianopoli, F.
Cordaro, P. Fr\`e, L. Gualtieri, \emph{Non-Semisimple Gaugings of
D=5 N=8 Supergravity and FDA.s}, Class.Quant.Grav. 18 (2001)
395-414, hep-th/0009048; L. Andrianopoli, F. Cordaro, P. Fr\`e, L.
Gualtieri, \emph{Non-Semisimple Gaugings of D=5 N=8 Supergravity},
Fortsch.Phys. 49 (2001) 511-518, hep-th/0012203.
%
\bibitem{fluxgauge1}C. Angelantonj, S. Ferrara and M. Trigiante, JHEP
\textbf{0310}, 015, (2003), Phys. Lett. {\bf B582} (2004) 263
%
\bibitem{fluxgauge2}B. de Wit, H. Samtleben and M. Trigiante Phys. Lett. {\bf B583} (2004)
338 [hep-th/0311224]
%
\bibitem{fluxgauge3}C. Angelantonj, R. D'Auria, S. Ferrara and M.
Trigiante Phys. Rev. Letters {\bf B583} (2004) 331
%
\bibitem{lledolauramario} L. Andrianopoli, M.A. {Lled\'o}, M.
Trigiante \emph{The Scherk Schwarz mechanism as a flux
compactification with internal torsion} hep-th/0502083.
\bibitem{cremmerjulia} E. Cremmer and B. Julia \emph{Supergravity Theory in eleven
dimensions},
Phys Lett. \textbf{B76} (1978) 409, \emph{The SO(8) supergravity} Nucl. Phys. {\bf B159} (1979)
141.
\bibitem{comments}  P. Fr\'e, \emph{
Comments on the 6--index photon in D=11 supergravity and the gauging of free differential algebras},
  Class.\ Quant.\ Grav.\  {\bf 1} (1984) L81.
\bibitem{fredauria} R. D'Auria and P. Fr\'e, \emph{Geometric
supergravity in D=11 and its hidden supergroup} Nucl. Phys.
\textbf{B201} (1982) 101.
\bibitem{congiani} L. Castellani, P. Fr\'e, F. Giani, K. Pilch and P.
van Nieuwenhuizen \emph{Gauging of D=11 supergravity} Ann. Phys.
\textbf{146} (1983) 35.
\bibitem{sullivan} D. Sullivan \emph{Infinitesimal computations in
topology} Bull. de l'Institut des Hautes Etudes Scientifiques, Publ.
Math. \textbf{47} (1977)
\bibitem{castdauriafre} L. Castellani, R. D'Auria, P. Fr\'e
\emph{Supergravity and superstrings: a geometric perspective}, World
Scientific, Singapore 1991.
\bibitem{GuidoSergioFDA} G. Dall'Agata and S.
Ferrara \emph{Gauged supergravity algebras from twisted tori compactifications with fluxes}
arXiv:hep-th/0502066.
\bibitem{RiccaMarioSergioFDA1} R. D'Auria, S. Ferrara, M. Trigiante,
\emph{$E_{7(7)}$ symmetry and dual gauge algebra of M-theory on a
twisted seven torus}
arXiv:hep-th/0504108.
\bibitem{RiccaGuidoSergioFDA} G. Dall'Agata, R. D'Auria and S.
Ferrara \emph{Compactifications on twisted tori with fluxes and free differential algebras}
arXiv:hep-th/0503122.
\bibitem{curvatureFDA1} R. D'Auria, S. Ferrara, M. Trigiante,
{\emph{Curvatures and potential of M-theory in D=4 with fluxes and
twist}} arXiv:hep-th/0507225.
\bibitem{spagnoli1} I.A. Bandos, J.A. de Azcarraga, J.M. Izquierdo,
M. Picon, O. Varela, \emph{On the underlying gauge group structure of
D=11 supergravity} Phys. Lett. {\bf B596} (2004) 145 [hep-th/0406020]
\bibitem{spagnoli2}  I.A. Bandos, J.A. de Azcarraga,
M. Picon, O. Varela, \emph{On the formulation of D=11 Supergravity and the composite nature of its three form field}
Ann. of Physics \textbf{317} (2005) 238 [hep/th0409100]
\bibitem{nastase} H. Nastase \emph{ Towards a Chern-Simons M-theory
of $\mathrm{Osp(1|32)} \times \mathrm{Osp(1|32)}$} hep-th/0306269
\bibitem{horava} P. Horava, \emph{M-theory as a holographic field
theory} Phys. Rev. {\bf D59} (1999) 04004 [hep-th/9712130]
\bibitem{leonardus} L. Castellani \emph{Lie derivative along
antisymmetric tensors and the M-theory superalgebra} hep-th/0508213
\end{thebibliography}
\end{document}